\documentclass[%
 reprint,
 amsmath,amssymb,
 aps,
 prb,
floatfix,
twocolumn
]{revtex4-2}

\usepackage{amsmath}
\usepackage{amssymb}
\usepackage{gensymb}
\usepackage{esint}
\usepackage{braket}
\usepackage{graphicx}
\usepackage{multirow}
\usepackage{lipsum} 
\usepackage{hyperref}
\usepackage[T1]{fontenc}
\usepackage[utf8]{inputenc}
\usepackage{chemfig}
\usepackage{tikz}
\usetikzlibrary{calc}
\usetikzlibrary{backgrounds}
\usetikzlibrary{fit}
\usetikzlibrary{positioning}
\usetikzlibrary{shapes.arrows}
\usetikzlibrary{arrows, decorations.pathmorphing, decorations.pathreplacing}
\usepackage{textcomp} % for \textdegree
\newcommand*\diff{\mathop{}\!\mathrm{d}}
\newcommand{\etal}{\textit{et al.}}
\newcommand{\LAMMPS}{\textsc{lammps}}
\newcommand{\drprobe}{\textsc{DrProbe}}
\newcommand{\fftw}{\textsc{fftw3}}
\newcommand{\fs}{\,\rm{fs}}
\newcommand{\ps}{\,\rm{ps}}

\newcommand{\thz}{\,\rm{THz}}
\newcommand{\mev}{\,\rm{meV}}

\usepackage{xcolor}
\hypersetup{
    colorlinks,
    linkcolor={blue!80!black},
    citecolor={blue!80!black},
    urlcolor={blue!80!black}
}

\definecolor{notecolor}{RGB}{0,0,191}
\definecolor{warningcolor}{RGB}{191,0,0}
\definecolor{donecolor}{RGB}{0,127,0}
\definecolor{todocolor}{RGB}{191,100,0}

\newlength{\minuslength}
\usepackage[sort&compress]{natbib}
\begin{document}

%---------- below are propositions from Jan and from Ángel ----------%
%\title{FTE-TACAW: A method for simulations of time-resolved vibrational electron energy loss spectra}
%\title{Combining Frozen Trajectory Excitation and Time Autocorrelation of Auxiliary Wavefunction for \textit{in silico} Time-Resolved Phonon Spectroscopy}
\title{Combining Frozen Trajectory Excitation and TACAW for \textit{in silico} Time-Resolved Vibrational Electron Energy Loss/Gain Spectroscopy}

\author{Wojciech Marciniak}%
	\email[email: ]{wojciech.marciniak@put.poznan.pl}%
    \affiliation{Department of Physics and Astronomy, Uppsala University, Box 516, 75120 Uppsala, Sweden}%
	\affiliation{Institute of Physics, Poznan University of Technology, Piotrowo 3, 60-965 Pozna\'n, Poland}%
\author{Joanna Marciniak}%
    \affiliation{Department of Physics and Astronomy, Uppsala University, Box 516, 75120 Uppsala, Sweden}%
	\affiliation{Institute of Molecular Physics, Polish Academy of Sciences,  M. Smoluchowskiego 17, 60-179 Pozna\'n, Poland}%
\author{Jos\'e \'Angel Castellanos-Reyes}%
	\affiliation{Department of Physics and Astronomy, Uppsala University, Box 516, 75120 Uppsala, Sweden}%
\author{J\'an Rusz}%
	\affiliation{Department of Physics and Astronomy, Uppsala University, Box 516, 75120 Uppsala, Sweden}%

\date{\today}% It is always \today, today,

%##############################%
%########## ABSTRACT ##########%
%##############################%

\begin{abstract}
Seeing that ultrafast (picosecond timescale) vibrational electron energy loss spectroscopy (EELS) should soon be experimentally realizable, we present \textit{in silico} approach capable of providing insight from the computational physics perspective.
We present a framework that combines frozen trajectory excitation (FTE) with time auto-correlation of auxiliary wavefunctions (TACAW) to study the time-dependent spectral response of non-equilibrium lattice dynamics in a way comparable directly to experiment\,\textemdash{}\,(scanning) transmission electron microscope EELS, (S)TEM-EELS. 
In this approach, a selected phonon excitation is first introduced into an equilibrium molecular dynamics trajectory using FTE, after which the atomic positions during subsequent relaxations are treated with short-time TACAW analysis performed at different pump\textendash{}probe delays. 
This yields momentum- and energy-resolved electron-scattering signals bearing a phonon imprint during the relaxation process, going beyond time-dependent diffuse-scattering intensities alone. 
We demonstrate the approach for fcc-Ni and 3C-SiC and discuss the observed phonon mode coupling and spectral redistribution during phonon relaxation.

\end{abstract}

%Use showkeys class option if keyword
\maketitle

%\tableofcontents

%##################################%
%########## INTRODUCTION ##########%
%##################################%

\section{\label{sec:intro}Introduction}

%---------- a bit of introduction to the introduction ----------%

Ultrafast electron diffraction (UED), as the experimental technique, is an analog of time-resolved crystallography that uses electrons as a probe rather than X-rays \cite{filippetto_ultrafast_2022}.
UED uses a pump\textendash{}probe setup.
There, a short pulse excites the system out of equilibrium (the ``pump''). 
Afterward, another short pulse passes through the sample (the ``probe''), and, ideally, a direct electron detector counts the electrons that got scattered by the sample as a function of scattering angle. 
 
In a typical experiment, the measured signal is associated with a finite delay time $\Delta t$ after excitation and a finite acquisition time $\tau$ determined by the probe duration. 
By scanning the delay, one obtains a time series of diffraction patterns that tracks the structural response of the system at time scales that can reach the attosecond regime.
Because electrons couple strongly to matter, UED offers invaluable insights into the dynamics of atomic~\cite{qi_breaking_2020} and electronic motion~\cite{hui_attosecond_2024}.
In the present work, we focus on lattice dynamics in the femtosecond-to-picosecond regime.

In practice, UED is usually performed in stroboscopic mode, since the signal-to-noise ratio (SNR) of a single acquisition is typically insufficient. 
The experiment is therefore repeated many times for fixed $\Delta t$ and $\tau$, and the resulting diffraction patterns are accumulated. 
Repeating the procedure for different delays yields a diffuse-scattering signal that contains information about the evolution of vibrational populations, parametrized by $\Delta t$ and $\tau$. 
However, when the detector records only the angular distribution of scattered electrons, much of the microscopic information is obscured by integrating over energy.

%---------- EELS + advancements -- experiment ----------%

A complementary route is to introduce the energy resolution by letting the electron pass through a spectrometer\,\textemdash{}\,yielding the electron energy loss (or gain) spectra (EELS/EEGS)~\cite{carbone_dynamics_2009,li_time_2014,zheng_ultrafast_2020,feist_quantum_2015,houdellier_development_2018}.
Time-resolved EELS (TR-EELS) has reached sub-picosecond temporal resolution.
Still, the energy spread of the probe pulse currently prevents reaching meV-scale energy resolution in EELS needed to resolve vibrational modes, as current beam optics typically lead to energy spreads of order 0.6~eV or larger~\cite{bucker_electron_2016,wang_capturing_2023}. 
The obstacle here is, however, not fundamental, and a 500~fs pulse could in principle be compatible with meV-scale energy precision.
A highly promising route is to use ultracold electron sources \cite{claessens_ultracold_2005}, which have a mean transfer energy in the meV range \cite{raadt_subpicosecond_2023,huijts_coming_2025} and reach energy spreads of around 10~meV \cite{karkare_ultracold_2020}.

%---------- Theoretical models for ultrafast dynamics ----------%

Alongside experimental developments, substantial progress has also been made in the theoretical modeling of ultrafast dynamics. 
A large part of the efforts has focused on the intricate interplay between the coupled subsystems\,\textemdash{}\,the crystal lattice, electrons, and spins.
Early descriptions based on the two-temperature model, considering the crystal lattice and electrons~\cite{anisimov_electron_1974,allen_theory_1987}.
Much attention has also been devoted to spin\textendash{}electron energy transfer~\cite{Battiato_superdiffusive_2010}.
The subsequent inclusion of all (spin\textendash{}electron\textendash{}lattice) degrees of freedom in pursuit of explaining the ultrafast demagnetization by a laser pulse led to the development of the three-temperature model, which has been further refined~\cite{Beaurepaire_ultrafast_1996,Koopmans_explaining_2010,Zahn_lattice_2021,Pankratova_heat-conserving_2022}.
At the same time, large-scale, atomistically resolved intra-reservoir (e.g., phonon\textendash{}phonon) energy redistribution remains comparatively less explored, though approaches built on model Hamiltonians exist~\cite{Maldonado_tracking_2020,Ritzmann_theory_2020}.

%---------- advancements -- calculationa ----------%

Advances in experimental methods and theoretical models were accompanied by improvements in algorithmic implementations and a rise in computational power.
From a paradigm perspective, libraries such as KOKKOS~\cite{trott_kokkos_2022} and PyTorch~\cite{paszke_pytorch_2019} enabled efficient, versatile hardware compatibility.
Simultaneously, machine-learned interatomic potentials (ML-IAPs or MLIPs) appeared and evolved~\cite{wang_machine_2024}.
A variety of solutions exist, ranging from parametrized potentials trained to a specific system like GAP~\cite{bartok_gausian_2010} or SNAP~\cite{thompson_spectral_2015} to purely universal ones\,\textemdash{}\,so-called foundational models\,\textemdash{}\,like MEGNet~\cite{chen_graph_2019} or ORB~\cite{neumann_orb_2024}.
They can reach density functional theory (DFT)-level accuracy at much reduced computational cost, enabling simulation of systems containing thousands to millions of atoms.
Those potentials are actively screened and benchmarked, e.g., by the MatBench project \cite{riebesell_framework_2025}.
Recently, foundational interatomic potentials were declared ready for explicit atomic simulations of phonons~\cite{loew_universal_2025}.

%---------- segway to the direct scope of the work ----------%

These developments in both experimental and computational capabilities motivate direct atomistic simulations of ultrafast vibrational phenomena. 
Recently, the frozen trajectory excitation (FTE) method introduced a practical way to impose excitation of a selected vibrational mode onto a molecular-dynamics trajectory \cite{marciniak_fte_2026}. 
In its original formulation, FTE was combined with a quantum-excitation-of-phonons (QEP)-type analysis to simulate time-dependent total, elastic, and diffuse-scattering signals. 
It enables a direct comparison with UED, but it does not allow for explicit energy resolution.

%---------- goal -- FTE/TACAW for introducing energy resolution ----------%

To recover spectral information, a natural candidate is the time auto-correlation of auxiliary wavefunctions (TACAW) method \cite{Castellanos-Reyes_dynamical_2025}. 
TACAW has recently been shown to provide momentum- and energy-resolved electron-scattering spectra from molecular-dynamics trajectories and has already been used in the context of nanoscale magnon spectroscopy \cite{kepaptsoglou_magnon_2025}. 
Ongoing implementations further aim at efficient hardware-accelerated workflows, as illustrated by recent and ongoing developments such as PySlice \cite{walker_pyslice_2026}.
We consider both FTE and TACAW to be expandable towards other quasiparticle excitations, such as magnons~\cite{marciniak_fte_2026,Castellanos-Reyes_dynamical_2025}.
TACAW has already been used for a coupled spin-lattice system~\cite{castellanos_theory_2025}.

%---------- goal -- testing the framework with specific excitations ----------%

Here, we combine FTE-based mode-selective excitation with TACAW-based spectral analysis into a unified FTE/TACAW framework for non-equilibrium lattice dynamics. 
While time-dependent diffuse scattering reveals the presence of relaxation processes, FTE/TACAW further identifies the momentum- and energy-resolved channels through which the vibrational spectral weight is redistributed.

From the theory perspective, the central question addressed in the present work is whether TACAW can be extended from equilibrium trajectories to the short-time analysis of non-equilibrium relaxation dynamics.
From the practical standpoint, we show the capabilities and limits of FTE/TACAW in the time-, momentum-, and energy-resolved analysis of phonon relaxation following a controlled excitation. 
We demonstrate the approach for two model systems: fcc-Ni, a prototypical metallic system frequently studied in ultrafast contexts~\cite{Beaurepaire_ultrafast_1996,Tauchert_polarized_2022}, and 3C-SiC, a binary material with two inequivalent atomic sublattices and a richer phonon structure. 
In this way, we explore how controlled FTE excitation and delayed TACAW analysis can reveal spectral redistribution pathways otherwise hidden in energy-integrated scattering alone.

%##########################################%
%########## CALCULATIONS DETAILS ##########%
%##########################################%

\section{Calculation Details}

%---------- section overview ----------%

We start by describing the calculation procedure in detail.
In Sec.~\ref{sec:method}, we describe the proposed FTE/TACAW framework and compare it with the previously described FTE approach, which we refer to from now on as FTE/QEP.
Then, in Sec.~\ref{sec:tacaw}, we bring up the topic of TACAW and investigate its applicability to analyzing the time-dependent auxiliary wavefunction from non-equilibrium MD trajectories.
We also propose solutions to mitigate uncertainties stemming from such an application of TACAW.
Finally, in Sec~\ref{sec:parameters}, we provide a brief description of the software and parameters used at all calculation steps. 

%---------- listing the details provided in the Appendix ----------%

\subsection{\label{sec:method}Method}

%---------- FTE/TACAW vs original FTE comparison -- original scheme ----------%

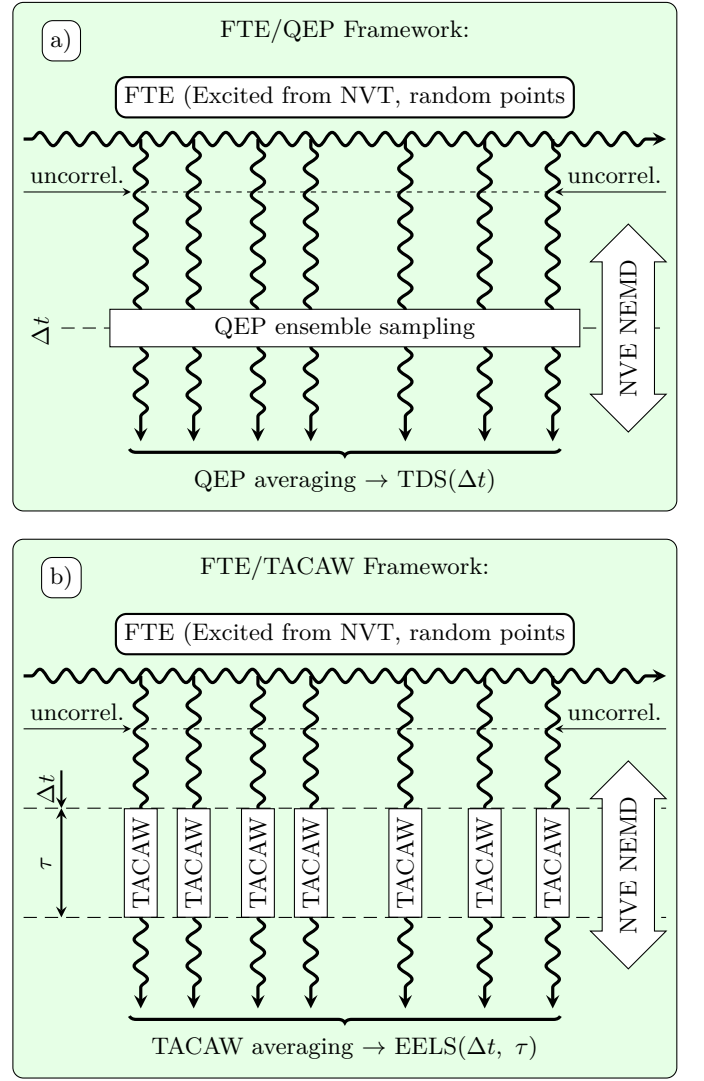
\begin{figure}

\begin{tikzpicture}[
    > = stealth,
    decoration={snake, pre length=3pt, post length=7pt},
    rounded/.style={rectangle, draw=black, rounded corners, align=center}, %minimum width=0.45\textwidth},
    rect/.style={rectangle, draw=black, align=center},
    empty/.style={align=center}]

    %---------- panel label ----------%
    \node[rounded, fill=white] at (0.5, 1.3) (label) {a)};

    %---------- original trajectory and FTE label ----------%
    \path[draw=black, ->, very thick, decorate] (0,0) -- node (fte-fte) [above, thick, yshift=2ex, rounded, fill=white] {FTE (Excited from NVT, random points} (8.5,0);
    \node[above = of fte-fte, yshift=-5ex] (txt-label) {FTE/QEP Framework:};

    %---------- QEP box ----------%
    \node[rect, fill=white, text width=6cm] at (4.25,-2.5) (qep) {QEP ensemble sampling};

    %---------- relaxation trajectories up to the QEP point ----------%
    \path[draw=black, very thick, decorate, decoration={snake, pre length=3pt, post length=0pt}] (1.55,0) -- (1.55,-2.26);
    \path[draw=black, very thick, decorate, decoration={snake, pre length=3pt, post length=0pt}] (2.25,0) -- (2.25,-2.26);
    \path[draw=black, very thick, decorate, decoration={snake, pre length=3pt, post length=0pt}] (3.10,0) -- (3.10,-2.26);
    \path[draw=black, very thick, decorate, decoration={snake, pre length=3pt, post length=0pt}] (3.80,0) -- (3.80,-2.26);
    \path[draw=black, very thick, decorate, decoration={snake, pre length=3pt, post length=0pt}] (5.05,0) -- (5.05,-2.26);
    \path[draw=black, very thick, decorate, decoration={snake, pre length=3pt, post length=0pt}] (6.10,0) -- (6.10,-2.26);
    \path[draw=black, very thick, decorate, decoration={snake, pre length=3pt, post length=0pt}] (7.00,0) -- (7.00,-2.26);

    %---------- comment (uncorrelated) ----------%
    \path[draw=black, dash pattern=on 2pt off 2pt] (1.55,-0.7) -- (7.0,-0.7);
    \path[draw=black, ->] (0,-0.7) -- node [above] {uncorrel.} (1.42,-0.7);
    \path[draw=black, <-] (7.03,-0.7) -- node [above] {\ uncorrel.} (8.5,-0.7);

    %---------- delta t, and corresponding dashed line ----------%
    \path[draw=black, dash pattern=on 5pt off 3pt] (7.4,-2.5) -- (8.5,-2.5);
    \path[draw=black, dash pattern=on 5pt off 3pt] (0.5,-2.5) node [left, rotate=90, xshift=2ex, yshift=2ex] {$\Delta t$} -- (1.14,-2.5);

    %---------- arrows showing relaxation ----------%
    \node[draw, fill=white, double arrow, minimum height=20mm, minimum width=10mm, double arrow head extend=1mm, anchor=center, rotate=90] at (8,-2.5) {NVE NEMD};

    %---------- relaxation trajectories continuation ----------%
    \path[draw=black, very thick, ->, decorate, decoration={snake, pre length=0pt, post length=5pt}] (1.55,-2.75) -- (1.55,-4);
    \path[draw=black, very thick, ->, decorate, decoration={snake, pre length=0pt, post length=5pt}] (2.25,-2.75) -- (2.25,-4);
    \path[draw=black, very thick, ->, decorate, decoration={snake, pre length=0pt, post length=5pt}] (3.10,-2.75) -- (3.10,-4);
    \path[draw=black, very thick, ->, decorate, decoration={snake, pre length=0pt, post length=5pt}] (3.80,-2.75) -- (3.80,-4);
    \path[draw=black, very thick, ->, decorate, decoration={snake, pre length=0pt, post length=5pt}] (5.05,-2.75) -- (5.05,-4);
    \path[draw=black, very thick, ->, decorate, decoration={snake, pre length=0pt, post length=5pt}] (6.10,-2.75) -- (6.10,-4);
    \path[draw=black, very thick, ->, decorate, decoration={snake, pre length=0pt, post length=5pt}] (7.00,-2.75) -- (7.00,-4);

    %---------- TACAW averaging curly brace ----------%
    \draw[decorate, very thick, decoration={brace, mirror, pre length=0pt, post length=0pt}] (1.4, -4.1) -- node (qep-averaging) [below, yshift=-1ex] {QEP averaging $\rightarrow$ TDS($\Delta t)$} (7.1, -4.1);

    %---------- box encompassing the whole panel ----------%
    \begin{scope}[on background layer]
        \node[rounded,fit=(txt-label)(qep-averaging), fill=green!10, style={minimum width=0.49\textwidth}] {};
    \end{scope}
    
\end{tikzpicture}

\vspace{1em}

\begin{tikzpicture}[
    > = stealth,
    decoration={snake, pre length=3pt, post length=7pt},
    rounded/.style={rectangle, draw=black, rounded corners, align=center}, %minimum width=0.45\textwidth},
    rect/.style={rectangle, draw=black, align=center, anchor=east, rotate=90},
    empty/.style={align=center}]

    %---------- panel label ----------%
    \node[rounded, fill=white] at (0.5, 1.3) (label) {b)};

    %---------- original trajectory and FTE label ----------%
    \path[draw=black, ->, very thick, decorate] (0,0) -- node (fte-tacaw) [above, thick, yshift=2ex, rounded, fill=white] {FTE (Excited from NVT, random points} (8.5,0);
    \node[above = of fte-tacaw, yshift=-5ex] (txt-label) {FTE/TACAW Framework:};

    %---------- TACAW boxes ----------%
    \node[rect, fill=white] at (1.55,-1.75) (tacaw1) {TACAW};
    \node[rect, fill=white] at (2.25,-1.75) (tacaw2) {TACAW};
    \node[rect, fill=white] at (3.10,-1.75) (tacaw3) {TACAW};
    \node[rect, fill=white] at (3.80,-1.75) (tacaw4) {TACAW};
    \node[rect, fill=white] at (5.05,-1.75) (tacaw5) {TACAW};
    \node[rect, fill=white] at (6.10,-1.75) (tacaw6) {TACAW};
    \node[rect, fill=white] at (7.00,-1.75) (tacaw7) {TACAW};

    %---------- relaxation trajectories up to the TACAW window ----------%
    \path[draw=black, very thick, decorate, decoration={snake, pre length=3pt, post length=0pt}] (1.55,0) -- (tacaw1);
    \path[draw=black, very thick, decorate, decoration={snake, pre length=3pt, post length=0pt}] (2.25,0) -- (tacaw2);
    \path[draw=black, very thick, decorate, decoration={snake, pre length=3pt, post length=0pt}] (3.10,0) -- (tacaw3);
    \path[draw=black, very thick, decorate, decoration={snake, pre length=3pt, post length=0pt}] (3.80,0) -- (tacaw4);
    \path[draw=black, very thick, decorate, decoration={snake, pre length=3pt, post length=0pt}] (5.05,0) -- (tacaw5);
    \path[draw=black, very thick, decorate, decoration={snake, pre length=3pt, post length=0pt}] (6.10,0) -- (tacaw6);
    \path[draw=black, very thick, decorate, decoration={snake, pre length=3pt, post length=0pt}] (7.00,0) -- (tacaw7);

    %---------- comment (uncorrelated) ----------%
    \path[draw=black, dash pattern=on 2pt off 2pt] (1.55,-0.7) -- (7.0,-0.7);
    \path[draw=black, ->] (0,-0.7) -- node [above] {uncorrel.} (1.42,-0.7);
    \path[draw=black, <-] (7.03,-0.7) -- node [above] {\ uncorrel.} (8.5,-0.7);

    %---------- delta t, tau, and corresponding dashed lines ----------%
    \path[draw=black, dash pattern=on 5pt off 3pt] (0,-1.75) -- (8.5,-1.75);
    \path[draw=black, dash pattern=on 5pt off 3pt] (0,-3.195) -- (8.5,-3.195);
    \path[draw=black, <->, thick] (0.5,-1.75) -- node [xshift=-1.5ex, rotate=90] {$\tau$} (0.5,-3.195);
    \path[draw=black, ->, thick] (0.5,-1.25) -- node [xshift=-1.5ex, rotate=90] {$\Delta t$} (0.5,-1.75);

    %---------- arrows showing relaxation ----------%
    \node[draw, fill=white, double arrow, minimum height=20mm, minimum width=10mm, double arrow head extend=1mm, anchor=center, rotate=90] at (8,-2.5) {NVE NEMD};

    %---------- relaxation trajectories continuation ----------%
    \path[draw=black, very thick, ->, decorate, decoration={snake, pre length=0pt, post length=5pt}] (tacaw1) -- (1.55,-4.4);
    \path[draw=black, very thick, ->, decorate, decoration={snake, pre length=0pt, post length=5pt}] (tacaw2) -- (2.25,-4.4);
    \path[draw=black, very thick, ->, decorate, decoration={snake, pre length=0pt, post length=5pt}] (tacaw3) -- (3.10,-4.4);
    \path[draw=black, very thick, ->, decorate, decoration={snake, pre length=0pt, post length=5pt}] (tacaw4) -- (3.80,-4.4);
    \path[draw=black, very thick, ->, decorate, decoration={snake, pre length=0pt, post length=5pt}] (tacaw5) -- (5.05,-4.4);
    \path[draw=black, very thick, ->, decorate, decoration={snake, pre length=0pt, post length=5pt}] (tacaw6) -- (6.10,-4.4);
    \path[draw=black, very thick, ->, decorate, decoration={snake, pre length=0pt, post length=5pt}] (tacaw7) -- (7.00,-4.4);

    %---------- TACAW averaging curly brace ----------%
    \draw[decorate, very thick, decoration={brace, mirror, pre length=0pt, post length=0pt}] (1.4, -4.5) -- node (tacaw-averaging) [below, yshift=-1ex] {TACAW averaging $\rightarrow$ EELS($\Delta t,\ \tau$)} (7.1, -4.5);

    %---------- box encompassing the whole panel ----------%
    \begin{scope}[on background layer]
        \node[rounded,fit=(txt-label)(tacaw-averaging), fill=green!10, style={minimum width=0.49\textwidth}] {};
    \end{scope}
    
\end{tikzpicture}
\caption{\label{fig:fte-fte_fte-tacaw}Comparison between the original FTE/QEP workflow, which yields time-dependent diffuse-scattering intensities, and the present FTE/TACAW workflow, which retains explicit energy resolution by performing TACAW on short trajectory segments extracted at a given pump\textendash{}probe delay. The parameters $\Delta t$ and $\tau$ denote, respectively, the time elapsed since the initial excitation and the temporal width of the local spectral probe.}
\end{figure}

To simulate time-dependent diffuse scattering from non-equilibrium molecular-dynamics (NEMD) trajectories, Ref.~\cite{marciniak_fte_2026} introduced the frozen trajectory excitation (FTE) method combined with an accompanying analysis based on the QEP formalism~\cite{forbes_quantum_2010,lugg_atomic_2015}.
In that framework, one starts from an equilibrium trajectory $\{\vec{R}_i,\vec{v}_i\}$ of a computational system and modifies it in reciprocal space so as to reflect the overpopulation of a selected region of phonon phase space. 
The resulting excited trajectory $\{\vec{R}_i',\vec{v}_i'\}$ is then sampled to generate a set of uncorrelated starting points for non-equilibrium relaxation runs that are subsequently done in the microcanonical ($NVE$) ensemble.
At a given delay time $\Delta t$ in those relaxation runs, one collects an ensemble of atomic configurations and performs multislice simulations for each of them, replicating the frozen phonon multislice (FPMS) approach~\cite{loane_thermal_1991}.
QEP-based ensemble averaging of the resulting exit wavefunctions $\phi$ yields the total and vibrational scattering intensities as functions of delay time $\Delta t$ and in-plane momentum transfer $\vec{q}_\perp$:
\begin{align}
    \label{eqn:qep-1} I_{\rm incoh}(\vec{q}_{\perp}, \Delta t) &= \frac{1}{N_S}\sum\limits_{i=1}^{N_S} \lvert\phi(\vec{q}_{\perp}, \vec{R}_i(\Delta t))\rvert^2 \\
    \label{eqn:qep-2} I_{\rm coh}(\vec{q}_{\perp}, \Delta t) &= \left\lvert\frac{1}{N_S}\sum\limits_{i=1}^{N_S} \phi(\vec{q}_{\perp}, \vec{R}_i(\Delta t))\right\rvert^2 \\
    \label{eqn:qep-3} I_{\rm vib}(\vec{q}_{\perp}, \Delta t) &= I_{\rm incoh}(\vec{q}_{\perp}, \Delta t) - I_{\rm coh}(\vec{q}_{\perp}, \Delta t),
\end{align}
where $I_{\rm incoh}$, $I_{\rm coh}$, and $I_{\rm vib}$ represent respectively the total scattering intensity (corresponding to the incoherent average), the elastic intensity (coherent average), and the vibrational intensity (thermal diffuse scattering\,\textemdash{}\,TDS).
Such FTE/QEP procedure, schematically summarized in Fig.~\ref{fig:fte-fte_fte-tacaw}(a), is well suited for direct comparison with time-resolved diffuse-scattering experiments\,\textemdash{}\,the ultrafast electron diffraction (UED).

%---------- FTE/TACAW vs original FTE comparison -- whats new in this approach ----------%

That formulation, however, remains energy integrated, obscuring the underlying physical information. 
Disentangling the contributions of phonon modes at specific energies for each in-plane momentum transfer could significantly enhance understanding of the underlying physical phenomena.
TACAW provides exactly this type of spectral information in equilibrium trajectories \cite{Castellanos-Reyes_dynamical_2025}. 
Therefore, in this present work, we replace the QEP-type post-processing step with a delayed short-time TACAW analysis. 
Instead of evaluating only isolated snapshots at a given delay $\Delta t$, we extract a short trajectory segment of duration $\tau$ starting at that delay and compute the time-dependent auxiliary exit wavefunction along that segment. 
Fourier transforming the corresponding signal yields a momentum- and energy-resolved spectrum associated with that probe window. 
Averaging over sufficiently many independent relaxation trajectories then yields a time-resolved spectrum that explicitly depends on both delay and probe duration, and inherently conserves the time\textendash{}energy uncertainty principle.
The corresponding workflow\,\textemdash{}\,FTE/TACAW\,\textemdash{}\,is shown schematically in Fig.~\ref{fig:fte-fte_fte-tacaw}(b).
A key challenge with this approach is TACAW's reliance on equilibrium system dynamics at the outset.

\subsection{\label{sec:tacaw}TACAW in non-equilibrium systems}

%---------- TACAW for NEMD justification ----------%

The central methodological step in this work is the extension of TACAW from equilibrium molecular dynamics to short-time analysis of non-equilibrium relaxation trajectories.
To motivate this extension, we first recall the equilibrium TACAW picture \cite{Castellanos-Reyes_dynamical_2025}. 
Introducing the inelastically scattered wave $\psi_{n\rightarrow{m}}(\vec{q})$ corresponding to crystal system transition from state $\ket{n}$ to state $\ket{m}$, within the operator $\hat{\phi}(\vec{q})$ describing beam propagation:
\begin{equation}
    \psi_{n\rightarrow m}(\vec{q}) = \braket{m|\hat{\phi}(\vec{q})|n},
\end{equation}
the energy-resolved scattering signal can be written as
\begin{equation}
    \label{eq:TACAW-energy-resolution}
    I(\vec{q},E) = \sum_{n,m} \frac{e^{-\beta E_n}}{Z}
    \left| \braket{m|\hat{\phi}(\vec{q})|n} \right|^2
    \delta\!\left(E-(E_m-E_n)\right),
\end{equation}
where $\beta=(k_{\mathrm B}T)^{-1}$ and $Z$ is the partition function. 
This expression can be re-expressed as a Fourier transform of an equilibrium correlation function,
\begin{equation}
    I(\vec{q},E) = \int_{-\infty}^{\infty} \frac{\diff t}{2\pi\hbar}
    e^{-iEt/\hbar} c_{\phi\phi}(\vec{q},t),
\end{equation}
with
\begin{equation}
    c_{\phi\phi}(\vec{q},t) =
    \frac{1}{Z}\mathrm{Tr}\!\left[
    e^{-\beta\hat{H}}
    \hat{\phi}^{\dagger}(\vec{q},0)\hat{\phi}(\vec{q},t)
    \right].
\end{equation}

The key practical approximation in TACAW implementations is that the quantum-mechanical time dependence of the auxiliary wavefunction is replaced by its parametric dependence on the time-dependent atomic positions:
\begin{equation}
    \phi(\vec{q},t) \approx \phi(\vec{q},\{\vec{R}_i(t)\}),
    \label{eq:tacaw_approx}
\end{equation}
using a trajectory obtained from molecular dynamics. 
This approximation is closely related to the picture discussed by Niermann in terms of the scattering matrix \cite{Niermann_scattering_2019}. 
In the TACAW formulation, the resulting energy-resolved signal is proportional to the spectral power density of the time-dependent auxiliary exit wavefunction,
\begin{equation}
    I(\vec{q_\perp},E) \propto |\phi(\vec{q}_\perp,E)|^2.
\end{equation}
This interpretation is particularly useful because it makes clear that TACAW can be viewed directly as a \textit{spectral analysis} of an evolving exit wavefunction rather than only as a reformulation of an equilibrium Kubo-type correlation function.

That viewpoint provides a natural route to the non-equilibrium case considered here. 
For a probe window of duration $\tau$ starting at delay $\Delta t$, we evaluate the auxiliary wavefunction along a short segment of a non-equilibrium relaxation trajectory and Fourier transform it over that finite interval. 
Averaging the resulting spectral intensities over many independent relaxation trajectories yields the working expression
\begin{equation}
    I(\vec{q_\perp},E,\Delta t,\tau) \propto
    \left\langle
    |\phi(\vec{q_\perp},E,\Delta t,\tau)|^2
    \right\rangle,
    \label{eq:fte-tacaw}
\end{equation}
which is the central equation of the present FTE/TACAW approach. 
Here, the dependence on $\Delta t$ describes the time elapsed since the initial excitation, while $\tau$ specifies the temporal width of the local spectral probe.
The reasoning above does not depend on the quasiparticle encoded in $\phi(\vec{q},t)$.
%
%If the excitation method could be extended in the future, e.g., to magnons, the TACAW usage described here will be possible, though in practical applications, the desired energy resolution and probe temporal width will differ, e.g., for phonons and magnons.
Snapshots of relaxation trajectories for, e.g., magnon excitations~\cite{Pankratova_heat-conserving_2022}, can be treated by the same TACAW procedure outlined here.

An important caveat must be emphasized. 
In equilibrium TACAW, a harmonic quantum correction factor $\beta E/[1-e^{-\beta E}]$ is employed to restore detailed balance when classical molecular dynamics is used to approximate the underlying quantum correlation function. 
In the present non-equilibrium setting, however, the phonon populations both at the start of the simulation and during relaxation are mode-dependent, time-dependent, and generally non-thermal.
From another angle of view, the probability of the system being in state $\ket{n}$, expressed in Eq.~\ref{eq:TACAW-energy-resolution} through the thermal occupation, does not depend on a single, well-defined equilibrium temperature, and no equally simple detailed-balance correction is available.
The present FTE/TACAW formulation therefore introduces an additional approximation beyond Eq.~\eqref{eq:tacaw_approx}. 

To partially address the above caveat, we propose the following procedure 
%to restore detailed balance. 
of a non-equilibrium quantum correction, inspired by the Kubo correction formula.
We note that the energy-resolved scattering intensity is proportional to the mean-squared displacement of the modes at that energy, and that is classically proportional to temperature~\cite{zeiger_lessons_2023}. Therefore:
\begin{equation}
    \label{eq:kubo-2}
    \frac{T(\vec{q_\perp},\omega,\Delta t)}{T_{\rm eq}} = \frac{I(\vec{q}_\perp,\omega,\Delta t)}{I_{\rm eq}(\vec{q}_\perp,\omega)} 
    \implies 
    T(\vec{q}_\perp,\omega,t) = \frac{I(\vec{q}_\perp,\omega,\Delta t)}{I_{\rm eq}(\vec{q}_\perp,\omega)} T_{\rm eq}
\end{equation}
retrieves the effective classical phonon mode temperature, based on the not-Kubo-corrected scattering intensity in equilibrium $I_{\rm eq}(\vec{q}_\perp,\omega)$ and at a given time delay in the NEMD simulation $I(\vec{q}_\perp,\omega,\Delta t)$.
Then, with $\frac{I_{\rm eq}(\vec{q}_\perp,\omega)}{I(\vec{q}_\perp,\omega,\Delta t)} = \alpha(\vec{q}_\perp,\omega,\Delta t)$, we could introduce a $(\vec{q},\omega)$-dependent hamonic quantum correction factor to the intensity at each time delay $\Delta t$:
\begin{equation}
    \label{eq:kubo-1}
    \frac{\beta E}{1 - e^{-\beta E}} 
    \rightarrow 
    \frac{\beta(\vec{q}_\perp,\omega,\Delta t)E}{1 - e^{-\beta(\vec{q}_\perp,\omega,\Delta t)}}
    = 
    \frac
    {\alpha(\vec{q}_\perp,\omega,\Delta t)\beta_{\rm eq}E}
    {1 - e^{-\alpha_(\vec{q}_\perp,\omega,\Delta t)\beta_{\rm eq}E}}
\end{equation}
This $\alpha(\vec{q}_\perp,\omega,\Delta t)$ factor incurs a small computational cost in the post-processing of every TACAW probe window.

Such an approach does not account for multi- and multiple-phonon scattering and further assumes equal lifetimes for any degenerate phonon modes under excitation.
Moreover, the phonon population (and thus the effective mode-dependent temperature) may change non-negligibly during the probe window $\tau$, while we assign a single temperature value to it. Nevertheless, this approximation becomes more accurate for shorter $\tau$, see Sec.~\ref{sec:fte-tacaw-resolution-limits} below.
Despite the mentioned shortcomings, it is expected to be a more accurate treatment than ignoring any correction in lieu of the Kubo factor altogether or assuming a single averaged temperature for all modes.
%
%Here, we focus primarily on the transient spectral redistribution and on the relative evolution of spectral features, rather than on claiming fully corrected absolute non-equilibrium spectral weights.
%
Future refinements of the theory should address this aspect more accurately.

\subsection{\label{sec:parameters}Parameters}

In this section, we summarize the key methodological choices and physical parameters relevant for interpreting the results. 
Additional technical details required for full reproducibility are provided in Appendix~\ref{app:details}.

\subsubsection{Software}

%---------- software ----------%

Both equilibrium and relaxation molecular-dynamics simulations were performed using the \LAMMPS{} simulation package~\cite{thompson_lammps_2022}. 
To introduce thermal overpopulation of specific phonon modes, the equilibrium trajectories were modified using the in-house FTE implementation described in Ref.~\cite{marciniak_fte_2026}. 
TEM exit wavefunctions, both in equilibrium and along the relaxation trajectories, were simulated in the multislice framework as implemented in \drprobe{}~\cite{barthel_dr_2018} and in \verb+pyms+~\cite{brown_python_2020}. 
In our calculations, we could benefit from GPU acceleration when using \LAMMPS{} and pyms~\cite{trott_kokkos_2022,paszke_pytorch_2019}.

\subsubsection{MD parameters in all simulations}

%---------- universal MD parameters ----------%

Molecular dynamics for fcc-Ni was performed using the SNAP interatomic potential~\cite{thompson_spectral_2015} in the parametrization of Zuo~\etal{}~\cite{zuo_performance_2020}. 
For 3C-SiC we employed the Vashishta interatomic potential~\cite{vashishta_interaction_2007}. 
All simulations were performed with a timestep of 1\fs{}.
Atomic positions and velocities sampled from the canonical ($NVT$) and isothermal\textendash{}isobaric ($NpT$) ensembles were generated using the Shinoda equations of motion \cite{shinoda_rapid_2004}, based on the Nos\'e\textendash{}Hoover chain modification of the Parrinello\textendash{}Rahman thermostat and barostat~\cite{parrinello_polymorphic_1981}. 
%WM: I'd say it's best suited here.
%Integrators used closely follow time-reversible measure-preserving Verlet and rRESPA integrators by Tuckerman~\etal{}~\cite{tuckerman_liouvilleoperator_2006,martyna_constant_1994}
The corresponding integrators closely follow the time-reversible measure-preserving Verlet and rRESPA schemes of Tuckerman~\etal{}~\cite{tuckerman_liouvilleoperator_2006,martyna_constant_1994}.
The non-equilibrium relaxation trajectories were subsequently generated in the microcanonical ($NVE$) ensemble using a plain time integration with the velocity-Verlet algorithm.

\subsubsection{Structures}

%---------- Models -- structures and supercells ----------%

\begin{figure}[h!]
\includegraphics[width=0.45\columnwidth]{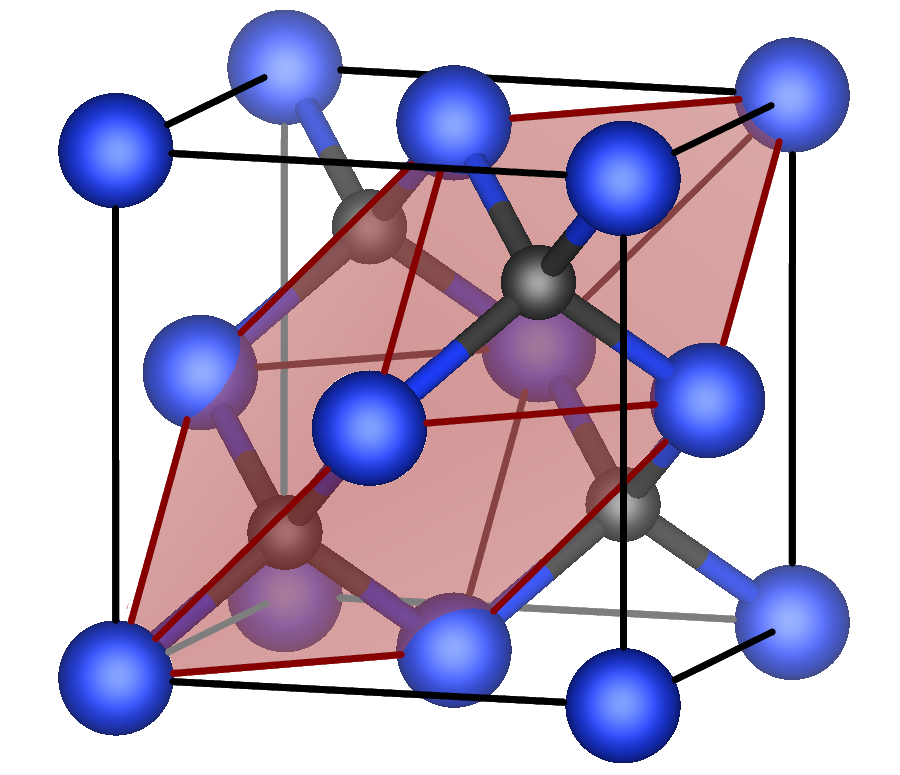}\includegraphics[width=0.45\columnwidth]{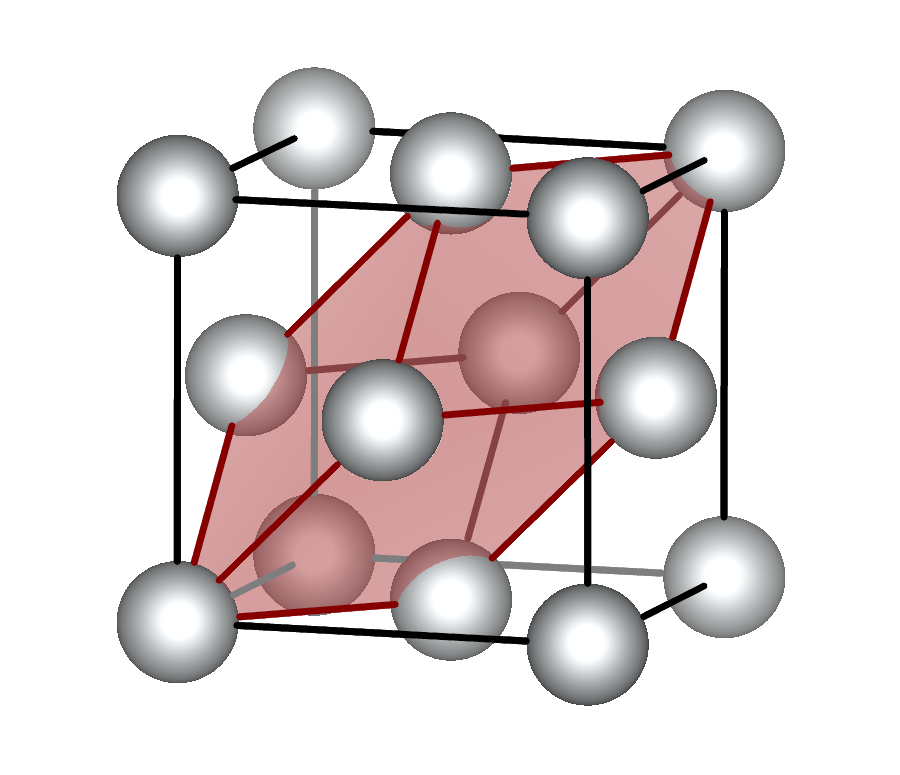}
\caption{\label{fig:unit-cells}Conventional and primitive unit cells used for 3C-SiC (left) and fcc-Ni (right). The black lines indicate the conventional cubic cells, while the shaded rhombohedra indicate the primitive cells. The structural difference between monatomic fcc-Ni and binary 3C-SiC is central to the later discussion of branch-selective and sublattice-sensitive redistribution.}
\end{figure}

Both systems considered here belong to the face-centered-cubic Bravais lattice. 
The first is fcc-Ni, a metallic system of Cu prototype with space group 225 ($Fm\bar{3}m$). 
The second is 3C-SiC, a binary zinc-blende structure with space group 216 ($F\bar{4}3m$), consisting of two inequivalent sublattices occupied by atoms of strongly different mass. 
For most molecular-dynamics calculations, we use the conventional cubic unit cells, containing four atoms for fcc-Ni and eight atoms for 3C-SiC, respectively, as illustrated in Fig.~\ref{fig:unit-cells}.
To plot the phonon dispersions, separate calculations were performed using primitive cells for both the MD simulation and the phonon structure calculation.
We did it to avoid visual artifacts from possible discontinuities that could arise due to remapping the supercell onto primitive cells to unfold the spectral features into the primitive Brillouin zone.

From these unit cells, supercells were constructed to reach sample dimensions of approximately $10\times10\times10$~nm$^3$, thereby obtaining sufficient resolution of the spatial wavelengths associated with the relevant phonon modes. 
This procedure yielded a $28\times28\times28$ supercell for fcc-Ni and a $24\times24\times24$ supercell for 3C-SiC when using the experimental lattice constants. 
Periodic boundary conditions were imposed in all three spatial directions.

\subsubsection{Equilibrium}

%---------- equilibrium trajectory preparation ----------%

Both systems were first equilibrated in the $NpT$ ensemble at 300~K under zero external pressure while preserving cubic symmetry and with cubic periodic boundary conditions. 
The resulting equilibrium lattice parameters, 3.5329~\AA{} for fcc-Ni and 4.37~\AA{} for 3C-SiC, were then used throughout the remaining calculations, yielding total system sizes of $(9.89212\,\mathrm{nm})^3$ and $(10.488\,\mathrm{nm})^3$, respectively. 
A higher number of decimal points in fcc-Ni when compared to SiC reflects much quicker stabilization of volume fluctuations in $NpT$ simulations.
From the equilibrated structures, five independent 10-ps long equilibrium trajectories at 300~K were generated. 
Snapshots $\{\vec{R}_i,\vec{v}_i\}$ were stored every 10\fs{} for 3C-SiC and every 50\fs{} for fcc-Ni in order to resolve all phonon branches of interest.

\subsubsection{Excitation}

%---------- excitations ----------%

Excitations were performed via multiplicative filters indexed by primitive unit cell repetitions, as described in Ref.~\cite{marciniak_fte_2026}.
To avoid discontinuities in time after transforming the excited trajectory back to the time domain, we used a Tukey (cosine-tapered) window \cite{bloomfield_fourier_2004}, with tapering over the first and last 5\% of the trajectory. 
While the FTE can introduce an arbitrary excitation pattern, here we introduced three narrow artificial FTE excitations localized in the $(\vec{q},\omega)$-space chosen deliberately to probe particular interesting cases:

\begin{itemize}
    \item[\textbf{Ni-$\Delta$}] in fcc-Ni: along the $\Delta$ ($\Gamma$\textendash{}X) high-symmetry line, we excite a point on the transverse acoustic (TA) branch such that the $(2\vec{q}_{\rm ex},2\omega_{\rm ex})$ point falls on the longitudinal-acoustic (LA) branch,
    \item[\textbf{SiC-$\Delta$}] in 3C-SiC: along the $\Delta$ ($\Gamma$\textendash{}X) high-symmetry line, we excite a point on the approximately linear portion of the TA branch such that $(2\vec{q}_{\rm ex},2\omega)_{\rm ex}$ and $(3\vec{q}_{\rm ex},3\omega_{\rm ex})$ remain on the same branch,
    \item[\textbf{SiC-$\Sigma$}] in 3C-SiC: along the $\Sigma$ ($\Gamma$\textendash{}K\textendash{}X) high symmetry line, we excite a point on the LA branch such that the $(2\vec{q}_{\rm ex},2\omega_{\rm ex})$ point lies on a transverse-optical (TO) branch.
\end{itemize}

All excitations used the same general filter shape, modified relative to Ref.~\cite{marciniak_fte_2026}. 
In particular, the filters are rod-like, being narrower in the $q_x,q_y$-plane and elongated along $q_z$, and the excitation is not imposed symmetrically in $\vec{q}$.
Instead of symmetry in both $\vec{q}$ and $\omega$, here we require symmetry only in $(\vec{q},\omega)$ vs $(-\vec{q},-\omega)$, i.e., the minimal condition guaranteeing real-valued atomic displacements.
The exact filter parameters are listed in Appendix~\ref{app:details}. 
In all cases, the applied excitation increased the system temperature by less than 3\%, from 300~K to at most 309~K.

\subsubsection{Relaxation}

%---------- relaxation trajectory preparation ----------%

Because of the Tukey window applied during the excitation procedure in $(\vec{q},\omega)$ space, we limited the portion of the transformed trajectory used for further analysis.
After the back-transform, we use only the part of the trajectory where the Tukey window function equals 1.
Consequently, snapshots $\{\vec{R}_i',\vec{v}_i'\}$ from the first and last 0.5\ps{} were excluded from being starting points for subsequent relaxation runs. 
From each excited trajectory, five snapshots were selected around 1, 3, 5, 7, and 9\ps{} from the beginning of the trajectory.
Each actual snapshot was chosen randomly within a $\pm100$~fs interval around the nominal time in order to reduce residual correlations. 
These snapshots served as initial conditions for subsequent 30-ps $NVE$ relaxation trajectories. 
In total, this yielded 25 relaxation trajectories for each excitation case. 
The sampling rates were the same as for the equilibrium trajectories, namely 50\fs{} for fcc-Ni and 10\fs{} for 3C-SiC.

\subsubsection{TEM propagation and spectroscopy}

%---------- multislice parameters ----------%

For each relaxation snapshot $\{\vec{R}_i\}(\Delta t)$, multislice simulations were performed with plane-wave illumination at 60~keV. 
We used \verb+pyms+~\cite{brown_python_2020} to compute the configuration-dependent exit wavefunction $\phi(\vec{q}_{\perp},\vec{R}_i(\Delta t))$, while \drprobe{}~\cite{barthel_dr_2018} was used to verify selected results.
For the FTE/QEP results, we performed suitable averaging with an external Python script.

%---------- TACAW parameters ----------%

In the FTE/TACAW workflow, the same auxiliary exit-wavefunction data were additionally processed in the time domain. 
We used a total TACAW sampling interval of 1.5\ps{}, consisting of a 1\ps{} signal window followed by 0.5\ps{} of zero padding. 
Prior to the Fourier transform, such time series $\phi(\vec{q}_{\perp},\vec{R}_i(\Delta t))$ was multiplied by a windowing function centered at half the signal length (0.5\ps{}) and with the same width as the data (1\ps{}).
We used a Hann window, corresponding to a maximally tapered Tukey window, in order to reduce spectral leakage and improve dynamic range.

%############################################%
%########## RESULTS AND DISCUSSION ##########%
%############################################%

\section{Results and Discussion}

In the following section, we demonstrate the practical application of the FTE/TACAW framework.
We first establish the equilibrium vibrational reference for the two systems through displacement\textendash{}displacement autocorrelation and identify the selected excitation points in reciprocal space in subsection~\ref{sec:initial}. 
We then analyze the subsequent relaxation on two complementary levels. 
In subsection~\ref{sec:fte-qep}, we examine time-dependent intensities integrated over a small aperture in FTE/QEP diffraction patterns, which provide the closest analog to ultrafast diffuse-scattering observables and bridge the work to Ref.~\cite{marciniak_fte_2026}. 
In subsection~\ref{sec:fte-tacaw}, we inspect selected transient momentum\textendash{}energy maps obtained from FTE/TACAW ($I(\vec{q}_{\perp},\omega)$), which reveal how the spectral weight evolves along selected reciprocal-space cuts. 
We also follow the time evolution of 2-dimensional data cuts from the 4-dimensional dataset provided by FTE/TACAW\,\textemdash{}\,$I(\vec{q}_{\perp_x},\vec{q}_{\perp_y},\omega,\Delta t)$\,\textemdash{}\,in order to separate the redistribution of spectral weight from possible peak shifts or broadening.
Finally, in Sec.~\ref{sec:fte-tacaw-resolution-limits} we investigate the limits of FTE/TACAW time resolution when applied to one selected excitation, namely SiC-$\Sigma$.
We show how much further benefits one could get from going to the sub-picosecond timescale.

%#####################################%
%########## INITIAL RESULTS ##########%
%#####################################%

\subsection{\label{sec:initial}Initial results}

%---------- comment on method to get $S(\vec{q},\omega)$ ----------%

Excitation is performed on autocorrelated molecular fields of displacements and velocities, as described in Ref.~\cite{marciniak_fte_2026}.
Hence, the first intermediate result is the magnitude of displacement\textendash{}displacement autocorrelation function that is proportional to the velocity\textendash{}velocity autocorrelation function and to the dynamical structure factor $S(\vec{q},\omega)$~\cite{lee_initio_1993,carreras_dynaphopy_2017}.
Pronounced local maxima of this intensity across the Brillouin zone trace the phonon dispersion.
Comparing this with experimental and quantum-mechanical-level density functional theory (DFT) data can yield valuable information to validate the use of the potentials employed in our work.

%---------------------------------------------------------%
%---------- POWER SPECTRUM: POTENTIALS ACCURACY ----------%
%---------------------------------------------------------%

\begin{figure}
\includegraphics{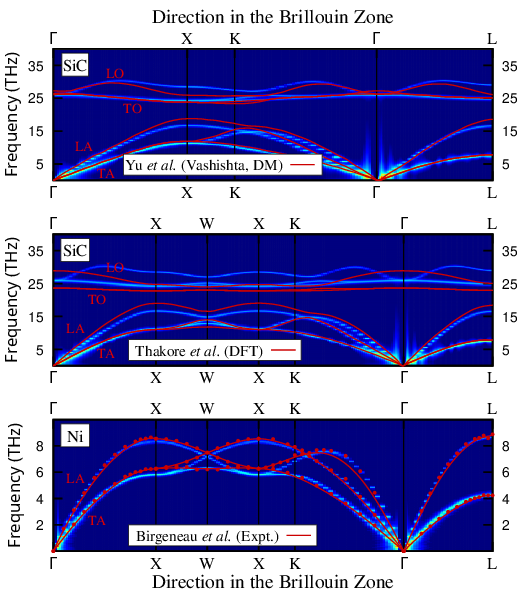}
\caption{\label{fig:equilibrium-dispersion-comparisons}Comparison of the equilibrium phonon dispersions used in the present work with literature references: dynamical-matrix calculations for SiC within the same Vashishta potential by Yu \etal{} (top), density-functional-theory results for SiC by Thakore \etal{} (middle), and experimental data for fcc-Ni by Birgeneau \etal{} (bottom). 
The LO (longitudinal optical), TO (transverse optical), LA (longitudinal acoustic), and TA (transverse acoustic) labels correspond to the respective phonon branches.
The figure provides the equilibrium reference for the branch geometry underlying the selected FTE excitation points and their intended transfer channels..}
\end{figure}

The equilibrium phonon reference is summarized in Fig.~\ref{fig:equilibrium-dispersion-comparisons}. 
This figure already serves two important purposes. 
First, it verifies that the interatomic potentials used in the MD simulations reproduce the overall branch structure of the target materials with sufficient fidelity for the present mode-selective study.
Second, it provides the reference against which the FTE excitation design should be understood. 

%---------- power spectrum: potentials accuracy -- SiC Vashishta in our simulations vs DM diagonalization by Yu ----------%

The top panel of Fig.~\ref{fig:equilibrium-dispersion-comparisons} presents a comparison of our SiC data to the density matrix diagonalization with the same (Vashishta) interatomic potential by Yu~\etal{}~\cite{yu_comparison_2024}.
We present the results along a path in the Brillouin Zone that is commensurate with their results.
Overall, the velocity\textendash{}velocity autocorrelation magnitude coincides well with the dispersion presented by Yu~\etal{}, especially for the transversal phonon branches, where there is close overlap.
For the longitudinal phonon branches, we observe noticeable deviation: acoustic phonon softening and flattening of the optical branch.
The effects are most pronounced at the Brillouin zone boundaries (X and L high-symmetry points), and we attribute this to our results being based on finite-temperature molecular dynamics simulations.

%---------- power spectrum: potentials accuracy -- SiC Vashishta in our simulations vs DFT by Thakore ----------%

Comparing with DFT data by Thakore~\etal{}~\cite{thakore_thermodynamic_2013}, presented in the middle panel of Fig.~\ref{fig:equilibrium-dispersion-comparisons}, presents a similar picture.
For the transverse modes, the quantitative agreement is good.
Especially in the TA branch, the overlap is exact, and the TO branch is reproduced up to a single THz.
Similar to the comparison to Yu~\etal{}~\cite{yu_comparison_2024}, there is a considerable LA mode softening, most prominent at the Brillouin Zone boundaries (X-W-X path and the L high-symmetry point).
LO mode is reproduced qualitatively well, though at an elevated frequency.
A noticeable difference occurs in the Brillouin Zone center ($\Gamma$), where both optical modes overlap, though DFT predicts a clear separation.
Overall, the comparison shows the expected separation between the acoustic and optical manifolds and reproduces the broad qualitative structure found in both earlier calculations and density-functional-theory data.

%---------- power spectrum: potentials accuracy -- Ni in our simulations vs experiment by Birgeneau ----------%

Bottom panel of Fig.~\ref{fig:equilibrium-dispersion-comparisons} presents a comparison of our velocity\textendash{}velocity autocorrelation magnitude in SNAP interatomic potential to the experimental measurements by Birgeneau~\etal{}~\cite{thakore_thermodynamic_2013}.
We use a simulation protocol very similar to that in Ref.~\cite{marciniak_fte_2026}, with a slightly smaller supercell.
The calculated branch structure agrees sufficiently well with the experimental acoustic phonon trends for the present proof-of-principle analysis. 
For our purposes, the most relevant point is not the sub-meV accuracy of every branch in both 3C-SiC and fcc-Ni systems, but rather that the chosen excitation and target locations lie in parts of the dispersion that are described consistently enough for the interpretation of energy redistribution pathways to remain meaningful.

%---------------------------------------------------------------------------------%
%---------- POWER SPECTRUM: SiC -- POWER SPECTRUM OF Si VS C VIBRATIONS ----------%
%---------------------------------------------------------------------------------%

\begin{figure}
\includegraphics{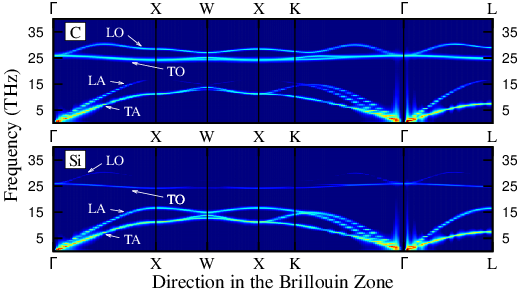}
\caption{\label{fig:dispersion-Si-C-contributions}Equilibrium spectral intensity in 3C-SiC resolved into contributions associated with the C-dominated (top) and Si-dominated (bottom) vibrational response. 
The figure highlights the distinct sublattice character of different spectral regions and provides context for interpreting acoustic-to-optical redistribution in the non-equilibrium SiC cases.
The LO (longitudinal optical), TO (transverse optical), LA (longitudinal acoustic), and TA (transverse acoustic) labels correspond to the respective phonon branches.}
\end{figure}

The sublattice-resolved spectrum of 3C-SiC is shown in Fig.~\ref{fig:dispersion-Si-C-contributions}. 
This figure contains more physical information than a simple intensity decomposition. 
It shows that the vibrational visibility of the two sublattices is not uniform across the spectrum. 
In particular, the acoustic and optical regions do not carry the same relative weight on the Si and C sublattices. 
We see a much stronger contribution to the total vibrational power spectrum from Si for the lower-frequency acoustic modes, especially LA.
Conversely, the contribution to the intensity of the optical branch comes mostly from C.
Such behaviour can be expected for two elements of considerably different atomic mass.
This matters for the later non-equilibrium analysis, especially for the SiC-$\Sigma$ case, where the intended transfer channel connects an acoustic excitation to an optical branch. 
In that situation, a transfer of spectral weight is not only a redistribution in energy and momentum; it may also reflect a change in the dominant sublattice character of the visible vibrational response.

%-------------------------------------------------------%
%---------- DIFFRACTOGRAMS AT THE EXCITATIONS ----------%
%-------------------------------------------------------%

\begin{figure*}
\includegraphics[width=0.98\textwidth]{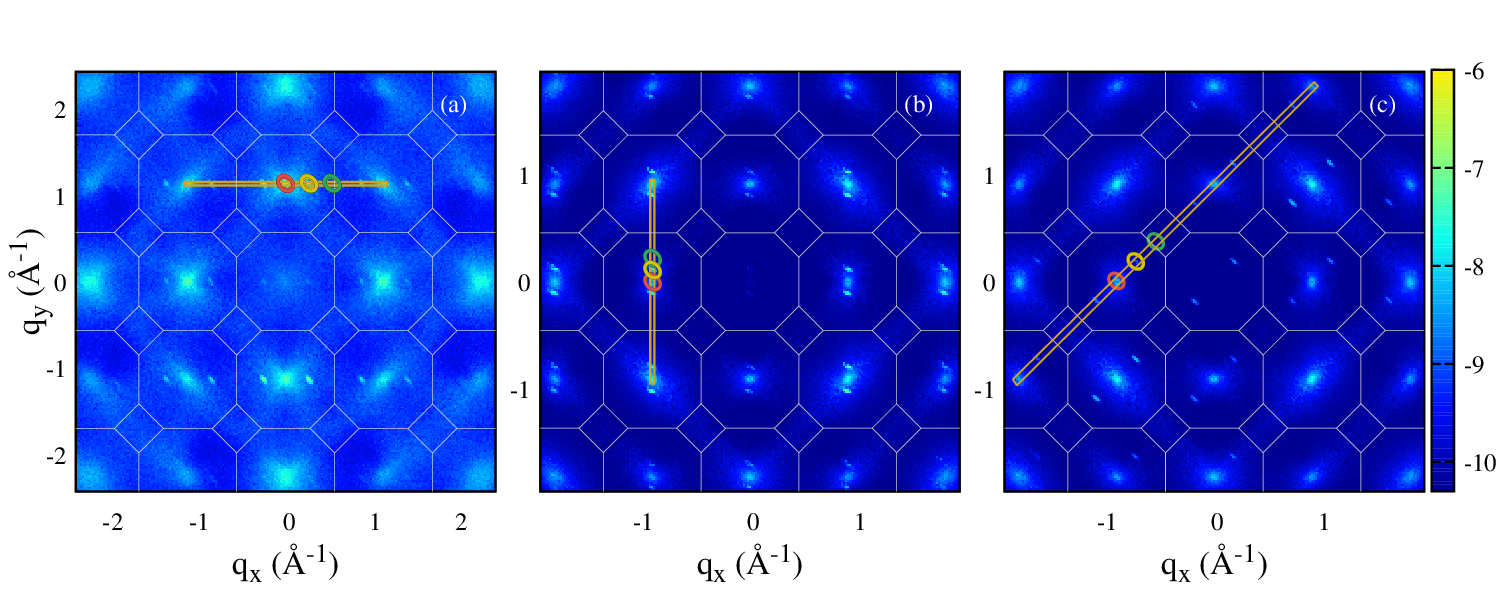}
\caption{\label{fig:excitation_0_frame}Thermal diffuse scattering at the time of the excitation for fcc-Ni (a) and 3C-SiC (b,c). 
Panels (b) and (c) correspond to the $\Gamma$\textendash{}X and $\Gamma$\textendash{}K\textendash{}X excitation geometries in SiC (SiC\textendash{}$\Delta$ and SiC\textendash{}$\Sigma$), respectively. 
The marked points identify the selected excitation location and the corresponding harmonic or transfer locations used later to monitor the energy redistribution. 
The finite reciprocal-space width of the excitation should be interpreted as a localized packet rather than a single infinitely sharp mode.}
\end{figure*}

The reciprocal-space localization of the imposed excitations is illustrated in Fig.~\ref{fig:excitation_0_frame}, which presents the electron diffraction patterns obtained via multislice simulations for system snapshots extracted from the excited trajectory. 
All diffraction patterns present geometry commensurate with the (001) zone axis of a cubic system.
The fcc Brilloin zone projections around every Bragg spot are marked with solid grey lines.  
The small integration apertures used in further time-dependent equilibration runs are marked with the green, yellow, and red ellipses.
The excitation is clearly visible on all panels.
Due to the more general excitation protocol, multi-phonon excitations at double momentum transfer (2$\vec{q}_{\perp{\rm ex}}$) are not as pronounced as in Ref.~\cite{marciniak_fte_2026}.
Only near the central Bragg spot, traces of the multi-phonon excitations can be found, most pronounced in Fig.~\ref{fig:excitation_0_frame}(b), though a faint spot is visible also in Fig.~\ref{fig:excitation_0_frame}(a).
No elevated intensity in 2$\vec{q}_{\perp{\rm ex}}$ in Fig.~\ref{fig:excitation_0_frame}(c) is visible, though it would be located outside of the first Brillouin zone.
The figure confirms that the FTE filters produce sharply localized non-equilibrium perturbations in reciprocal space, while still having finite widths. 
This point is important. 
The excitation is localized enough to justify discussing specific target transfer channels, but it is not infinitely narrow, meaning that the subsequent relaxation should be understood as the evolution of a packet in $(\vec{q},\omega)$ space rather than of a single perfectly isolated normal mode. 

%-------------------------------------%
%---------- SECTION SUMMARY ----------%
%-------------------------------------%

Taken together, Figs.~\ref{fig:equilibrium-dispersion-comparisons}\textendash{}\ref{fig:excitation_0_frame} define the reference before the relaxation. 
They present a well-defined starting point for the question about how a localized increase in vibrational mode population redistributes in momentum and energy during the following $NVE$ relaxation.

%###########################################%
%########## RELAXATION -- FTE/QEP ##########%
%###########################################%

\subsection{\label{sec:fte-qep}Relaxation\,\textemdash{}\,FTE/QEP}

%---------- point-like aperture UED intensity in time ----------%

\begin{figure}
\includegraphics{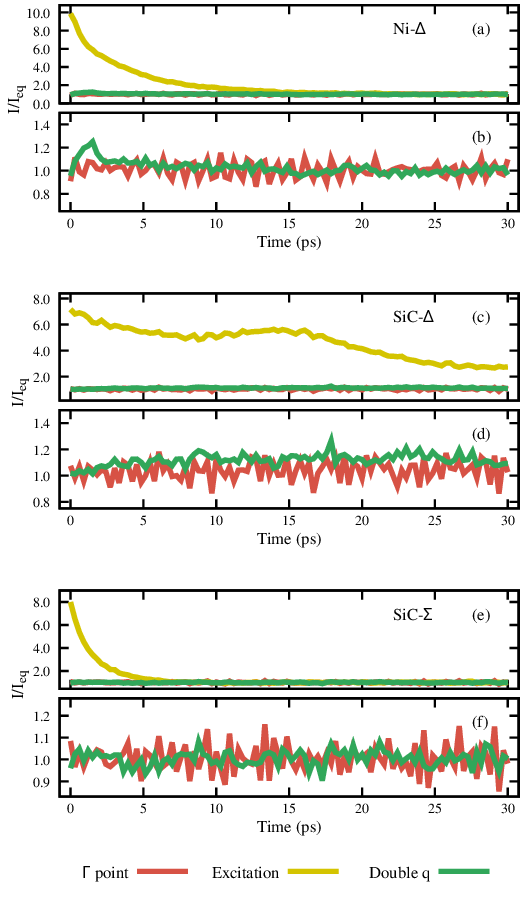}
\caption{\label{fig:I_q_vs_t}Time evolution of integrated scattering intensity at selected reciprocal-space regions for the three excitation cases. 
Panels (a,b) show the $\Gamma$\textendash{}X ($\Delta$) excitation in fcc-Ni, panels (c,d) show the $\Gamma$\textendash{}X ($\Delta$) excitation in 3C-SiC, and panels (e,f) show the $\Gamma$\textendash{}K\textendash{}X excitation in 3C-SiC. 
In each case, the monitored points include the zone center, the excited wavevector, and the corresponding harmonic or transfer point targeted by the excitation design. 
The figure shows that the strongest transient response occurs at the transfer-related momentum channel and that the relaxation phenomenology differs appreciably between Ni-$\Delta$, SiC-$\Delta$, and SiC-$\Sigma$ excitations.
The colors are commensurate with the apertures marked in Fig~\ref{fig:excitation_0_frame}.}
\end{figure}

The first level of information is obtained from delay-dependent intensities, shown in Fig.~\ref{fig:I_q_vs_t}, integrated over the small apertures introduced for the equilibrium picture. 
These quantities are the closest analogue of observables accessible in time-resolved diffuse scattering and therefore provide the natural baseline against which the additional value of the FTE/TACAW analysis should be judged. 
In each case, we monitor three representative momentum spots: the zone center ($\Gamma_{(0,2,0)}$), the original excitation point ($\Gamma_{(0,2,0)} + \vec{q}_{\rm ex}$), and a second point based on the excitation design, that is expected to capture effects related to potential energy transfer towards higher harmonic ($\Gamma_{(0,2,0)} + 2\vec{q}_{\rm ex}$).

In all three excitation scenarios, the largest deviations occur at the momentum transfer corresponding to the excitation rather than at $\Gamma$. 
Also, the relaxation behavior varies across cases.

%---------- I vs t in fcc-Ni ----------%

Relaxation process for the Ni-$\Delta$ case, presented in Fig.~\ref{fig:I_q_vs_t}(a\textendash{}b) fully shows a relatively fast exponential decay of the enhanced signal near the excitation point.
Over the first 0.5\ps{}, a section of faster decay can be observed, accompanied by an increase of intensity in double the momentum transfer, seen in Fig.~\ref{fig:I_q_vs_t}(b).
This clearly shows that an efficient energy transfer is occurring.
Subsequently, both the primary and secondary excitation relax further at a steadier rate.
Overall, it shows that the energy dissipation across phonon modes is non-trivial.
This general trend is similar to the excitation at $\frac{1}{2}\vec{q}_X$ in Ref~\cite{marciniak_fte_2026}.
The relaxation observed in this work is faster, which may be related to the fact that the excitation place in the Brillouin zone has been engineered precisely so that $(2\vec{q}_{\rm ex},2\omega_{\rm ex})$ lies exactly on another phonon branch.
As we will show in the next subsection, this is the point at which the main energy transfer occurs.
Full relaxation occurs in the span of 15\ps{}, consistent with the timescales of fcc-Ni lattice response reported by Pankratova~\etal{}~\cite{Pankratova_heat-conserving_2022}.

%---------- I vs t in SiC -- along the Delta path ----------%

In contrast, relaxation in the SiC-$\Delta$ case, presented in Fig.~\ref{fig:I_q_vs_t}(c\textendash{}d), appears more gradual and less monotonic, suggesting a more distributed or multi-step intensity redistribution pathway.
The excitation itself is low-frequency\,\textemdash{}\,3.5\thz{}.
Over 30\ps{}, the excitation is far from being relaxed.
This timescale is consistent with phonon\textendash{}phonon scattering rates in 3C-SiC calculated from second- and third-order force constants obtained in DFT and density functional perturbation theory (DFPT) by Wang~\etal{}~\cite{wang_capturing_2023}.
The scattering rate of $10^{-2}\thz{}$ corresponds to the phonon lifetime in the order of 100\ps{}.
The probability of intra-mode energy transfer is low, but as we can see in Fig.~\ref{fig:I_q_vs_t}(d), the energy transfer towards $2\vec{q}_{\rm ex}$ remains the main source of the relaxation.
However, long-time-scale additional oscillations in intensity appear, resembling a beat with another process of similar oscillatory character.
It suggests that the energy transfer to $(2\vec{q}_{\rm ex},2\omega_{\rm ex})$ is rivaled by another energy-dissipation mechanism that occurs in other regions of the q-space than the three areas we have focused on.

%---------- I vs t in SiC -- along the Sigma path ----------%

The SiC-$\Sigma$ case, presented in Fig.~\ref{fig:I_q_vs_t}(e\textendash{}f), exhibits an initial response that is similarly strong but even more impulsive than in the Ni-$\Delta$ case, followed by rapid decay. 
The time scale is well below 10\ps{}, commensurate with the scattering rates above $10^{-1}\thz{}$ for the high-frequency range of the LA mode calculated by Wang~\etal{}~\cite{wang_capturing_2023}.
Such a finding also indicates efficient energy transfer, though we cannot pinpoint the exact mechanism from the UED diffraction patterns alone.
It is exactly the kind of information that is enabled by the FTE/TACAW analysis presented in the further section.

%---------- Overall results ----------%

All relaxation profiles contain small-scale oscillations associated with the dominant phonon mode frequency in the aperture.
Those oscillations are inherently related to the finite ensemble sampling and diminish with the square of the sample count~\cite{marciniak_fte_2026}.
Additionally, on the timescale shown across all panels, the intensity around the $\Gamma$ point remains comparatively weakly perturbed in all cases, exhibiting finite-time (THz-scale) oscillations resembling the ones that could be observed experimentally, for example, in Bi in Ref.~\cite{qi_breaking_2020}.
It indicates that the dominant early-time dynamics is not simply a uniform heating of the lattice but rather a redistribution among selected finite-momentum vibrational channels.

These observations already suggest a physically meaningful distinction between the three cases. 
In Ni-$\Delta$, the selected excitation appears to couple efficiently to double $\vec{q}_{\rm ex}$\,\textemdash{}\,green line in panel (b). 
In SiC-$\Delta$, the slower relaxation suggests a redistribution that remains to a greater extent within a related branch manifold, though the main energy transfer seems to happen also towards double $\vec{q}_{\rm ex}$\,\textemdash{}\,green line in panel (d)\,\textemdash{}\,though on a longer timescale and not as the main energy dissipation pathway.
No such clear dominant transfer could be identified for the SiC-$\Sigma$ case.
At this stage, we are not able to pinpoint the exact mechanisms, because integrated intensities alone do not reveal which spectral features carry the evolving weight. 
It is apparent, however, that the three excitation designs generate distinct relaxation phenomenology.

%#############################################%
%########## RELAXATION -- FTE/TACAW ##########%
%#############################################%

\subsection{\label{sec:fte-tacaw}Relaxation\,\textemdash{}\,FTE/TACAW}

%--------------------------------------------------------%
%---------- ILUSTRATION OF THE FOLLOWING PLOTS ----------%
%--------------------------------------------------------%

\begin{figure*}
\includegraphics[width=0.98\textwidth]{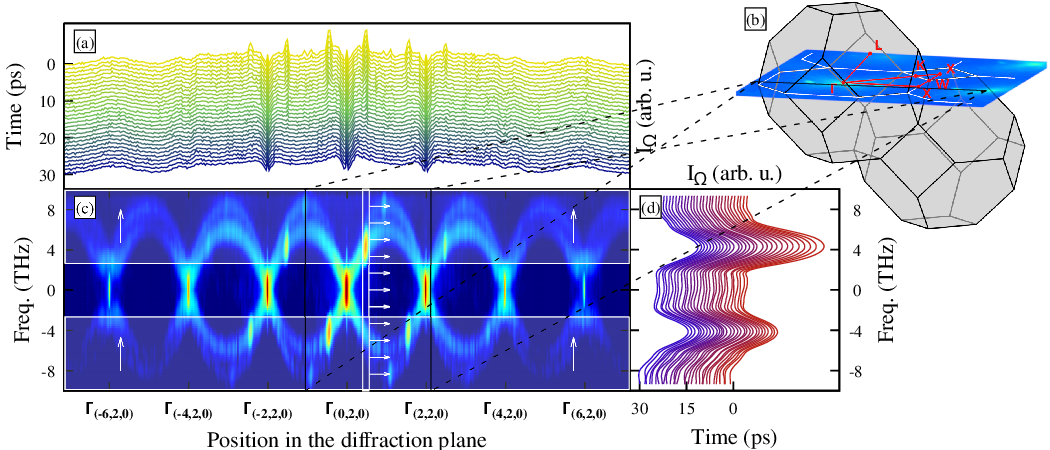}
\caption{\label{fig:BZ}An overview of the FTE/TACAW protocol and its capabilities.
Face-centered-cubic Brillouin zone and its cross-section by the projection plane used in the present diffraction geometry are presented in panel (b). 
The highlighted paths define relevant high-symmetry directions, including the reciprocal-space cuts employed for the analyses along $\Gamma$\textendash{}X ($\Delta$) and $\Gamma$\textendash{}K\textendash{}X ($\Sigma$) high-symmetry paths. 
The L point lies outside the diffraction plane, so from now on, it is excluded from the analysis.
Panel (c) presents the energy loss and gain fractional scattering intensity I($\vec{q}_\perp,\omega$) resolved along the $\Delta$ high symmetry line in fcc-Ni, including the harmonic quantum correction term.
Panels (a) and (d) show the post-processing capabilities of the 4D data resulting from FTE/TACAW.
Panel (a) presents the energy-filtered, $\vec{q}_\perp$-resolved scattering intensity.
Panel (d) presents the electron energy-loss and gain spectrum integrated over a small aperture centered on a specific momentum transfer $\vec{q}_\perp$ corresponding to the position of the excitation.
%
% --- point-like detector EELS(t).
}
\end{figure*}

The limitations of analysis offered by energy-integrated intensities shown in Fig.~\ref{fig:I_q_vs_t} are notable. 
Although the data establish that the selected momentum channels respond differently over time, it does not tell us whether the changing signal arises from a transfer between specific peaks, a broad redistribution across an entire branch, or actual shifts in positions of spectral features. 
This is precisely the reason to aim for introducing the energy resolution through the FTE/TACAW protocol.
In Fig.~\ref{fig:BZ}, we present the avenue for the diffraction pattern energy decomposition.

%---------- Brillouin Zone overlayed ----------%

Panel Fig.~\ref{fig:BZ}(b) shows the fcc Brillouin zone overlay over the example diffraction pattern.
High-symmetry lines measurable in the diffraction plane are shown in red.
The figure shows the justification for choosing the excitation points along the $\Delta$ and $\Sigma$ high-symmetry lines, as these lines pass through the $\Gamma$-point, allowing detection of both low- and high-frequency modes.
The third high-symmetry direction, passing through X\textendash{}W\textendash{}X high-symmetry points, is of less interest for us now, as it goes along the Brillouin Zone boundaries exclusively.
It also shows why, from this point, we do not consider the $\Lambda$ ($\Gamma$\textendash{}L) high symmetry line, as it is not oriented in the detection plane.

%---------- Additional energy coordinate -- spectrum ----------%

First of all, adding the energy resolution allows for clearer resolution of the excitation point as it evolves in time. 
Figure~\ref{fig:BZ}(c) presents the $\vec{q}_\perp$-resolved electron energy loss and gain spectrum for the fcc-Ni system snapshot during the relaxation process.
Strictly speaking, it is the double-sided TACAW power spectral density corrected with the non-equilibrium harmonic quantum correction term.
Two main observations can be made immediately.
First, the sharply defined excitation can be seen in $(\vec{q}_{\perp{\rm ex}},\omega_{\rm ex})$ and another spot appears in $(2\vec{q}_{\perp{\rm ex}},2\omega_{\rm ex})$.
Second, whatever process corresponding to energy loss can be observed in $(\vec{q}_\perp,\omega)$, a corresponding energy gain can be detected in $(-\vec{q}_\perp,-\omega)$.
This means the imposed excitation is not symmetric in $\vec{q}$, which makes the setup general and also means that symmetry-related redistribution should not be assumed automatically.
A more detailed comparison of the EELS/EEGS spectra along the relevant high-symmetry lines in both materials will follow shortly in the next subsection.

%---------- intensity integrated over energy, after Kubo factor and with frequency-filtering -- the same as intensity vs time, but with additional bonuses ----------%

Having access to the full spectrum provides clear advantages. 
First of all, electron diffraction can be recalculated including classical/quantum statistics correction in the form of the Kubo factor $\frac{\beta T}{1-e^{-\beta T}}$.
The results before integration back into the diffraction plane can also be filtered in energy, e.g., to remove the elastic scattering contribution to the total intensity.
An exemplary time-dependent intensity scan along the $\Delta$ high-symmetry line in fcc-Ni is presented in panel Fig.~\ref{fig:BZ}(a).

%---------- EELS and EEGS in time ----------%

Panel Fig.~\ref{fig:BZ}(d) presents another avenue\,\textemdash{}\,time-resolved EELS/EEGS spectrum in a detector.
In our case, it's a small aperture centered on a specific momentum transfer $\vec{q}_{\perp{\rm ex}}$ corresponding to either the position of the excitation or double the momentum transfer $2\vec{q}_{\perp{\rm ex}}$, as presented in Fig.~\ref{fig:excitation_0_frame}.

%%% JAN's NOTE: I read it until here.

%----------------------------------------------%
%---------- TACAW (spectrum) in time ----------%
%----------------------------------------------%

\begin{figure}
\includegraphics[width=0.99\columnwidth]{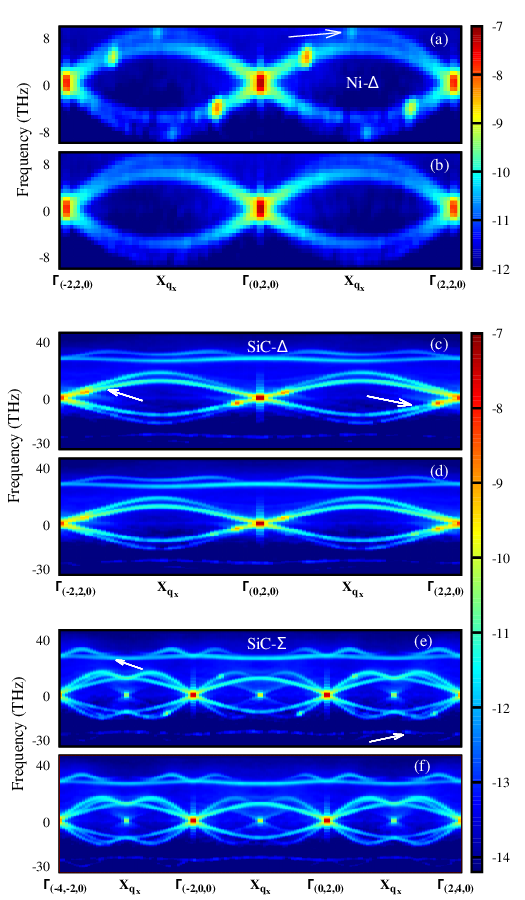}
\caption{\label{fig:spectrum-in-time}
Quantum-corrected EELS/EEGS spectra obtained with 1\ps{} temporal window starting at different time points during the relaxation.
Panel (a) presents the fcc-Ni $\Delta$ high-symetry path immediately after excitation.
The same path in the 29\textendash{}30 ps time window is presented in panel (b), showing a fully equilibrated spectrum.
Panel (c) presents the 3C-SiC $\Delta$ high-symmetry path immediately after excitation.
The same path in the time window if 20\textendash{}21\ps{} afterwards is shown in panel (d).
It is the time point when the energy transfer towards $(2\vec{q}_{\rm ex},2\omega_{\rm ex})$ is the most apparent.
In panel (e), we show the 3C-SiC $\Sigma$ high-symmetry path immediately after the excitation.
Panel (f) shows the same path during time window of 20\textendash{}21\ps{} after the excitation, showing a fully equilibrated structure.
The overall branch topology remains clearly visible across the relaxation, while the relative spectral weight is redistributed. 
This indicates that the non-equilibrium response is expressed mainly through redistribution of spectral weight rather than through a breakdown of the underlying branch structure.
}
\end{figure}

%---------- overall intro ----------%

In Fig.~\ref{fig:spectrum-in-time}, we present the EEL/EEG spectrum measured in the first picosecond after the excitation (a,c,e) \textit{versus} 20\ps{} after the excitation (b,d,f).
One immediate observation in all panels is that the equilibrium branch topology remains recognizable at both early and late times. 
The relaxation, therefore, does not create qualitatively new spectral structures. 
Instead, the dominant effect is a redistribution of spectral weight within the already available vibrational manifold. 
It indicates that, for the present perturbation strengths, the non-equilibrium lattice dynamics is best understood as a time-dependent repopulation of existing branches rather than as a large transient renormalization of the dispersion.

At the same time, the early- and late-time maps are not identical.
The excitation spot can be clearly seen in (a), (c), and (e) panels\,\textemdash{}\,for $\Delta t = 0$.
After 20\ps{}, as expected from diffraction intensity (Fig.~\ref{fig:I_q_vs_t}), the system is clearly equilibrated in case of fcc-Ni-$\Delta$ (b) and 3C-SiC-$\Sigma$ (f) excitations. 
In the case of 3C-SiC-$\Delta$, on the other hand, after 20 ps (Fig.~\ref{fig:spectrum-in-time}(d)), the excitation is still well visible, although the intensity has diminished.
In all cases, an interesting feature appears.
In the first TACAW time interval for Ni-$\Delta$ and SiC-$\Sigma$, and at a later time point in SiC-$\Delta$, a spot appears in double the momentum transfer and double the energy $(2\vec{q}_{\perp{\rm ex}},2\omega_{\rm ex})$ (marked with white arrows).
We identify it as the energy transfer towards $2\vec{q}_{\rm ex}$, suggested by both scattering intensity changes presented in Fig.~\ref{fig:I_q_vs_t}, and similar representation reported in the original article introducing FTE~\cite{marciniak_fte_2026}.

%---------- Ni-Delta ----------%

In Ni-$\Delta$ (Fig~\ref{fig:spectrum-in-time}(a)), a well-localized intensity peak at $(2\vec{q}_{\perp{\rm ex}},2\omega_{\rm ex})$ is clearly seen, appearing on a different acoustic branch.
Combined with the time-resolved electron diffraction pattern analysis, it indicates an energy transfer to $(2\vec{q}_{\rm ex},2\omega_{\rm ex})$\,\textemdash{}\,from TA to LA phonon branch\,\textemdash{}\,as the primary dissipation mechanism.
%
%---------- SiC-Delta ----------%
%
In the SiC-$\Delta$ geometry, The excitation was designed so that higher harmonics remain on the same (TA) phonon branch.
Probability of the energy transfer in such situation is low, leading to only a slight intensity rise at $(2\vec{q}_{\perp{\rm ex}},2\omega_{\rm ex})$, in Fig.~\ref{fig:spectrum-in-time}(d). 
This interpretation also agrees with the comparatively slow and less abrupt integrated-intensity decay seen in Fig.~\ref{fig:I_q_vs_t}.
Nevertheless, we identify such TA\textendash{}TA energy transfer as the main energy dissipation mechanism, as the slight spectral intensity increase at $(2\vec{q}_{\perp{\rm ex}},2\omega_{\rm ex})$ is the only feature we can confidently pinpoint.

%---------- SiC-Sigma ----------%

The corresponding spectra for the SiC-$\Sigma$ case are shown in Figs.~\ref{fig:spectrum-in-time}(e\textendash{}f). 
Here, the chosen excitation geometry was designed to place it on the LA branch so that $(2\vec{q}_{\rm ex},2\omega_{\rm ex})$ falls on an optical branch. 
The spectra again preserve the overall branch structure, but the redistribution pattern is more suggestive of transfer between manifolds of different character (acoustic\,$\rightarrow$\,optical). 
In particular, the comparison of early and late delay windows indicates that the intensity changes are not confined to acoustic phonons. 
Instead, the response again appears in $(2\vec{q}_{\perp{\rm ex}},2\omega_{\rm ex})$ and quickly spreads over a broader part of the spectrum, consistent with the idea that the relevant transfer does not depend on the involved phonon branches beyond vibrational state change probability.

%---------- summary and perspectives ----------%

The additional dimension of the data provides access to a range of insights that go beyond the scope of this work.
For example, as shown in Fig.~\ref{fig:dispersion-Si-C-contributions}, the optical branches in SiC are associated with the movement of C atoms, whereas the LA branch along the $\Sigma$ path is mainly associated with the movement of Si.
Hence, our results indicate an efficient Si\,$\rightarrow$\,C acoustic\,$\rightarrow$\,optical mode energy transfer.
The simple plane-wave TACAW used in this work does not allow further pursuit of this lead.
It does not, however, mean that this limitation cannot be lifted in further studies of particular systems.
For example, to obtain the signal resolved by a chemical element, displacements of any of the sublattices could be zeroed, followed by TACAW performed on such filtered trajectories. 
Alternatively, atomic-resolution scans could be performed, at the expense of dispersion, but would provide an energy-resolved picture of Si and C contributions.
Time resolution could be pushed further, below the picosecond acquisition times, depending on the system under investigation\,\textemdash{}\,an example of which we will show in Sec.~\ref{sec:fte-tacaw-resolution-limits} and in the Appendix~\ref{app:details}.
Direct phonon mode and sublattice decomposition, if achieved, could yield a different, complementary layer of information.
In combination with the energy resolution, such analysis allows observing features that would not be visible from the diffractogram intensity alone.

%----------------------------------------------------------------%
%---------- time integration -- intensity vs q in time ----------%
%----------------------------------------------------------------%

\begin{figure}
\includegraphics[width=0.9\columnwidth]{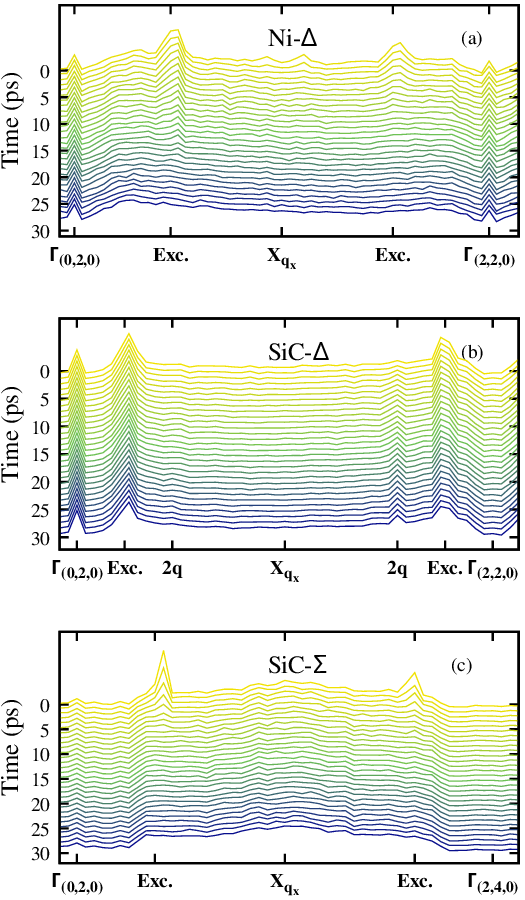}
\caption{\label{fig:I_q_vs_t_waterfall}Time evolution of the momentum-resolved energy-integrated intensity along the $\Gamma$\textendash{}X\textendash{}$\Gamma$ scan in fcc-Ni (a), along the $\Gamma$\textendash{}X\textendash{}$\Gamma$ scan in 3C-SiC (b), and along the $\Gamma$\textendash{}K\textendash{}X\textendash{}K\textendash{}$\Gamma$ scan in 3C-SiC (c).
The energy range around 0\mev{} is excluded from integration to avoid the elastic peaks.
The figure bridges the gap between the fully energy-integrated observables of Fig.~\ref{fig:I_q_vs_t} and the energy-resolved spectra of Fig.~\ref{fig:I_E_vs_t}. 
It shows how the momentum distribution evolves in time, while at the same time illustrating that an energy-integrated view alone obscures the spectral features that carry the changing weight.
}
\end{figure}

Figure~\ref{fig:I_q_vs_t_waterfall} provides an intermediate view between the fully integrated FTE/QEP traces and the fully energy-resolved FTE/TACAW spectra. 
The presented signal is obtained by integrating the FTE/TACAW spectra over selected energy ranges.
It directly corresponds to the time evolutions of the $\vec{q}_\perp$-space cut obtained by FTE/QEP, presented in Ref.~\cite{marciniak_fte_2026}.
However, we excluded from integration the $[-11,11]$~\mev{} ($[-2.67,2.67]$~\thz{}) energy range for the $\Delta$ direction in both systems and the $[41.36,41.36]$~\mev{} ($[-10,10]$~\thz{}) for the $\Sigma$ direction in 3C-SiC.
Such energy filtering demonstrates the advantage of TACAW analysis, which efficiently removes low-energy contributions from inelastic scattering at Bragg spots and from kinematically forbidden reflections.

%---------- Ni-Delta ----------%

The $\Gamma$\textendash{}X\textendash{}$\Gamma$ cut in fcc-Ni, presented in Fig.~\ref{fig:I_q_vs_t_waterfall}(a), shows how the momentum distribution of the integrated signal evolves in time. 
As the signal progresses in time, we see that two main features are exhibited: an intensity increase at the excitation point and at double the momentum transfer, which can be seen in the 0th frame and subsequently relax toward the equilibrium.
Although the increase in intensity near both recorded Bragg spots is similar, the timescale of the relaxation observed near $\Gamma_{(2,2,0)}$ is shorter than near $\Gamma_{(0,2,0)}$.
It may indicate that the excitation localizes closer to the phonon band center in time, detaching from the Ewald sphere cut through the Brillouin zone that is recorded in the diffraction plane.

%---------- SiC-Delta ----------%

Figure~\ref{fig:I_q_vs_t_waterfall}(b) presents the $\Gamma$\textendash{}X\textendash{}$\Gamma$ cut in 3C-SiC.
Here, the picture is different, yet consistent with Fig.~\ref{fig:I_q_vs_t}.
The intensity drop at the excitation peak is accompanied by an intensity increase at double the momentum transfer, which persists up to 30\ps{} into the relaxation.
The signal is asymmetric, showing more prominent features near $\Gamma_{(2,2,0)}$

%---------- SiC-Sigma ----------%

The $\Gamma$\textendash{}K\textendash{}X\textendash{}K\textendash{}$\Gamma$ cut in 3C-SiC is presented in Fig.~\ref{fig:I_q_vs_t_waterfall}(c).
There, although the energy range has been reduced to improve the contrast of the expected features, the energy transfer at twice the momentum transfer remains undetected in this way of visualizing the data.
Below, we will show how, with a different approach, the intensity appearing at $(2\vec{q}_{\perp{\rm ex}},2\omega_{\rm ex})$ can be unambiguously identified in the simulated data
%
%---------- summary for all panels ----------%
%
Overall, Fig.~\ref{fig:I_q_vs_t_waterfall} visualizes that there are no noticeable shifts in the positions of the peaks associated with the excitation and main momentum transfer. 

Energy filtering provides more information than the pure diffraction intensity reported in Ref.~\cite{marciniak_fte_2026}, revealing nontrivial changes in the momentum distribution.
Below, we show how the availability of energy resolution can further enhance our understanding of underlying excitation and relaxation processes and distinguish whether the observed changes arise from redistribution among separate spectral peaks or from broad intensity shifts within a single unresolved energy-integrated signal. 

%-------------------------------------------------------------------------%
%---------- time resolution -- point-like aperture EELS in time ----------%
%-------------------------------------------------------------------------%

%---------- figure ----------%

\begin{figure*}
\includegraphics[width=0.99\textwidth]{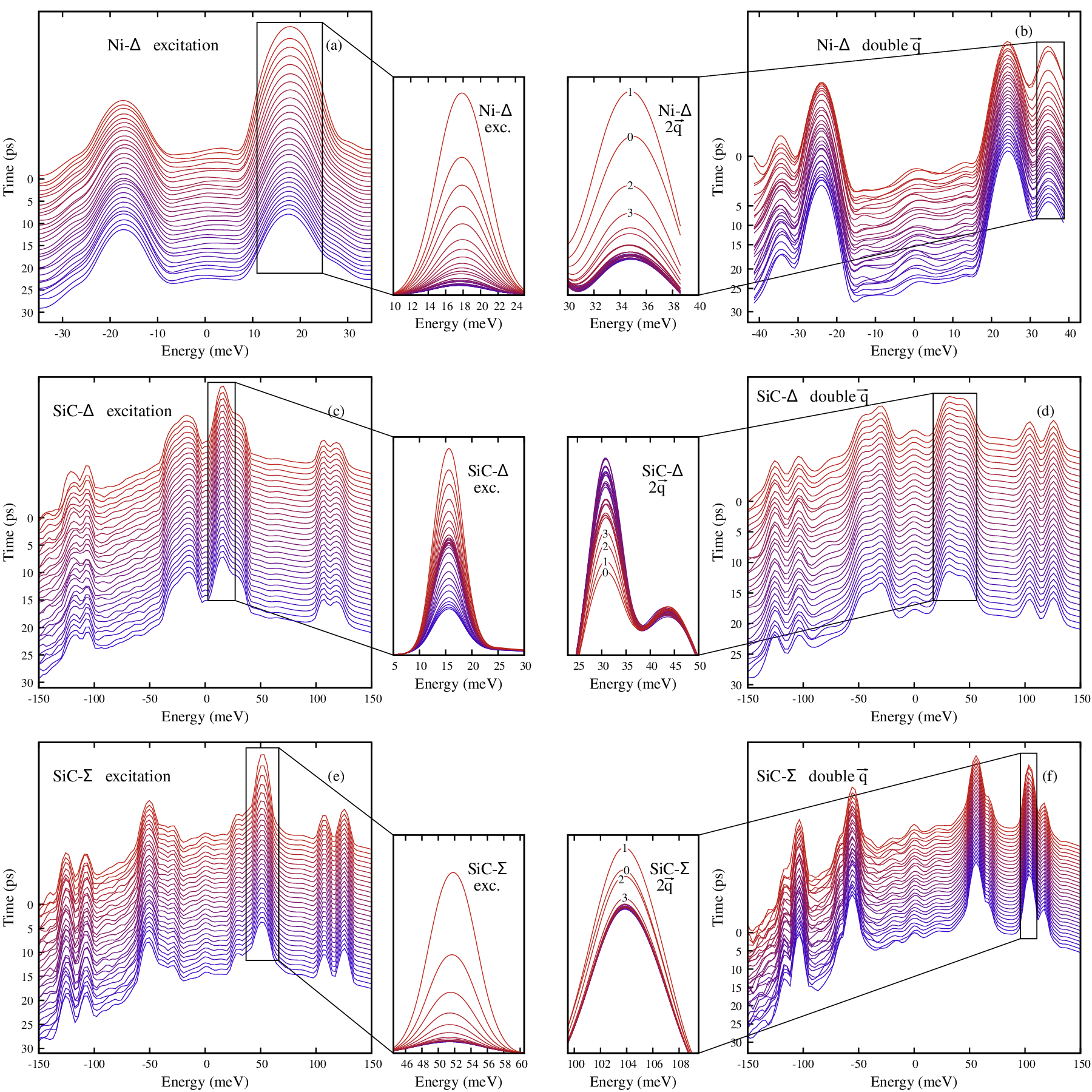}
\caption{\label{fig:I_E_vs_t}
Energy-resolved intensity evolution over time in the apertures presented in Fig.~\ref{fig:excitation_0_frame}, centered around the excitation (panels (a), (c), and (e)), and double the momentum transfer $2\vec{q}_{\perp{\rm ex}}$ (panels (b), (d), and (e)).
The electron energy loss and gain spectra near the excitation point and the corresponding doubled wavevector for the fcc-Ni $\Gamma$\textendash{}X (Ni-$\Delta$) excitation geometry are presented in panels (a) and (b), respectively.
Panels (c) and (d) present the EELS/EEGS spectra integrated over small apertures centered at the excitation in 3C-SiC along the $\Gamma$\textendash{}X path (SiC-$\Delta$) and the corresponding double the momentum transfer.
Panels (e) and (f) present the spectra integrated over corresponding apertures for the excitation in 3C-SiC along the $\Gamma$\textendash{}K\textendash{}X path (SiC-$\Sigma$).
All panels show that the relevant momentum points are spectrally distinct and that the doubled wavevector hosts the strongest and richest transient response. 
The dominant evolution is a redistribution of spectral weight among existing features rather than a large shift of the spectral positions themselves.
The underlying equilibrium spectra are easily discernible in all cases.
The accompanying inserts present certain features in more detail, excluding the waterfall plot style, to facilitate easier analysis.
}
\end{figure*}

Figure~\ref{fig:I_E_vs_t} presents the electron energy loss and gain spectra (EELS/EEGS).
We follow the energy-resolved intensity at the two most representative momentum points: the excitation point, and the doubled wavevector.
These plots reveal the FTE/TACAW framework's main advantage most clearly. 
The integrated traces of Fig.~\ref{fig:I_q_vs_t} and corrected, time-integrated traces in Fig.~\ref{fig:I_q_vs_t_waterfall} already showed that the excitation itself and the doubled wavevector carry the strongest transient signal.
Still, the signal-to-noise ratio achievable even with a decent statistic of 25 independent TACAW window averages has proven insufficient for SiC-$\Sigma$.
The energy-resolved plots, however, reveal how the signal is distributed across specific spectral features, yielding a more informative picture and a further increase in SNR where it's needed the most.

%---------- Ni-Delta ----------%

The Ni case is analyzed in more detail in Fig.~\ref{fig:I_E_vs_t}(a,b).
Here, the situation is the simplest.
As the excitation energy dissipates, the corresponding intensity\,\textemdash{}\,at around 20 meV in Fig.~\ref{fig:I_E_vs_t}(a)\,\textemdash{}\,decreases, and the underlying band-structure-dependent signal stabilizes.
Simultaneously, for the $(2\vec{q}_{\perp{\rm ex}},2\omega_{\rm ex})$\,\textemdash{}\,at around 40 meV in Fig.~\ref{fig:I_E_vs_t}(b)\,\textemdash{}\,we see an increase in the band intensity and a subsequent relaxation towards the equilibrium band structure.
Thanks to the added energy resolution, this effect is significantly more visible than in Fig.~\ref{fig:I_q_vs_t_waterfall}(a), where the signal is integrated over a wide energy range.

%---------- SiC-Delta ----------%

Fig.~\ref{fig:I_E_vs_t}(c,d) presents the time-dependent EELS/EEGS spectrum integrated around the SiC-$\Delta$ excitation and around the corresponding $2\vec{q}_{\perp{\rm ex}}$.
In both panels, we see a non-trivial energy transfer occurring.
It is associated with above-thermal-noise changes in most of the intensity peaks, corresponding to both acoustic and optical phonon bands.
The intensity drops non-monotonically in $(\vec{q}_{\perp{\rm ex}},\omega_{\rm ex})$ and intensity increases in $(2\vec{q}_{\perp{\rm ex}},2\omega_{\rm ex})$, indicating that the energy transfer between the corresponding points is the most pronounced.
It shows that the transfer-related channel is not represented by a single isolated intensity peak.

Instead, the internal balance of a bigger portion of the spectrum changes with the delay time.
The generation of higher harmonics, though, remains the main dissipation mechanism.

%---------- SiC-Sigma ----------%

In the SiC-$\Sigma$ case, presented in Fig.~\ref{fig:I_E_vs_t}(e,f), the situation is clearer.
In the excitation point, we observe a quick decay, even sharper than in Fig.~\ref{fig:I_q_vs_t_waterfall}.
The doubled wavevector case is more interesting because it appears to host multiple evolving features. 
The EELS/EEGS intensity in the range, where the excitation is located, changes beyond the thermal fluctuations
Nevertheless, the most pronounced feature is the clear energy transfer to $(2\vec{q}_{\rm ex},2\omega_{\rm ex})$, followed by rapid dissipation across the whole spectrum.
It finally explains the quick energy dissipation by efficient energy transfer from the Si acoustic mode to the optical mode, specifically TO.
Not only is the energy transfer from LA to the TO mode efficient, but the TO mode also dissipates energy efficiently.

%---------- overall ----------%

In all cases, the dominant evolution is again a change in spectral weight rather than a dramatic motion of peak positions, and the underlying band structure remains clearly visible. 
Any genuine peak motion, if present, is modest compared with the redistribution in amplitude.
The present data, therefore, suggest that the relaxation process is governed primarily by population redistribution among existing channels. 
This observation is important because it indicates that the main added value of the method, at least in the present examples, is not the detection of strong transient phonon softening or hardening, but rather the ability to resolve where the spectral weight flows in the joint momentum\textendash{}energy space during relaxation.

%----------------------------------------%
%---------- SUBSECTION SUMMARY ----------%
%----------------------------------------%

Taken together, Figs.~\ref{fig:BZ}\textendash{}\ref{fig:I_E_vs_t} show that the three excitation cases display genuinely different relaxation mechanics, with Ni-$\Delta$, SiC-$\Delta$, and SiC-$\Sigma$ showing different temporal and spectral redistribution patterns.
More importantly, the energy-resolved FTE/TACAW analysis reveals information about those patterns that is fundamentally inaccessible in integrated diffuse-scattering observables alone. 
Overall, the dominant non-equilibrium signature following the excitations in our work is not a large transient reshaping of the branch topology, but a time-dependent redistribution of spectral weight within the already existing vibrational manifold. 

%--------------------------------------------%
%---------- TIME RESOLUTION LIMITS ----------%
%--------------------------------------------%

\subsection{\label{sec:fte-tacaw-resolution-limits}Time-resolution limits of FTE/TACAW}

\begin{figure}
\includegraphics[width=0.9\columnwidth]{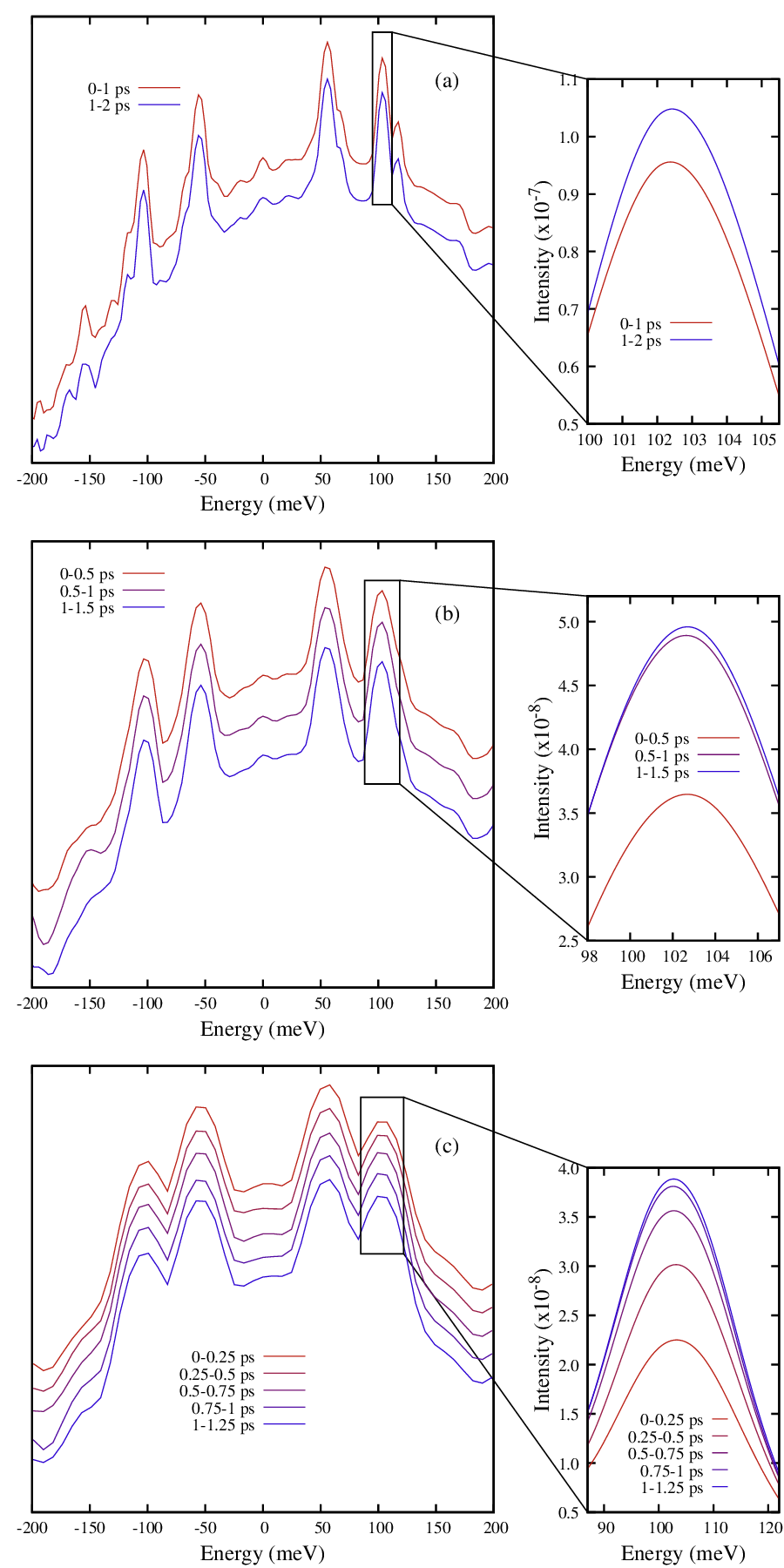}
\caption{\label{fig:EELS_vs_chunk}Waterfall plot of EELS/EEGS intensity near $2\vec{q}_{\perp{\rm ex}}$ in log scale using different measuring time window lengths with the same ratio of additional zero-padding.
Corresponding overlayed linear scale close-ups near $2\omega_{\rm ex}$ are presented in the insets to the right of each of the panels.
The plots show that energy resolution decreases as the time window length decreases.
Simultaneously, the increased time resolution enables more accurate analysis of the mode dynamics. 
}
\end{figure}

A natural question is how much time resolution can be achieved while keeping the important spectral features well resolved in energy.
The evolution of $(2\vec{q}_{\perp{\rm ex}},2\omega_{\rm ex})$ EELS/EEGS peak for the Ni-$\Delta$ and SiC-$\Sigma$ excitations, shown in Fig.~\ref{fig:I_E_vs_t}(b,f), indicates that higher time resolution could reveal significant additional information.
In the case of Ni-$\Delta$, increasing temporal resolution (which reduces energy resolution) is challenging because the excited phonon branch lies close to the corresponding $(2\vec{q}_{\perp{\rm ex}},2\omega_{\rm ex})$ and reduced energy resolution (accompanying the improved time resolution) would lead to a merging of the branches on the energy axis.
By contrast, for SiC-$\Sigma$, the acoustic branches that carry the excitation are well separated from the optical branches containing $(2\vec{q}_{\perp{\rm ex}},2\omega_{\rm ex})$.
It suggests a narrower $\tau$ window could be used.

Figure~\ref{fig:EELS_vs_chunk} shows a waterfall plot of EELS/EEGS intensity near $2\vec{q}_{\perp{\rm ex}}$ for different measurement window lengths in the SiC-$\Sigma$ excitation case.
We present spectra up to the one starting 1\ps{} after the excitation.
The (a\textendash{}c) panels present the spectrum on a logarithmic scale and the corresponding overlayed linear scale close-ups near $2\omega_{\rm ex}$ are presented in the insets to the right of each of the panels.

For a probe temporal width of 1\ps{} (See Fig.~\ref{fig:EELS_vs_chunk}(a)), the LA, LO, TO, and one of the non-degenerate TA modes are well resolved.
One cannot distinguish between the other transverse and longitudinal acoustic modes.
The reason here is partly the weaker electron scattering on certain modes, observed in both the measured and calculated spectra~\cite{Castellanos-Reyes_dynamical_2025,ohara_high_2023,plotkin_hybrid_2020}.
Excpectedly, the energy resolution decreases as the time window shortens, causing loss of transverse–longitudinal mode resolution below $\tau \approx 1\, {\rm ps}$.
For this excitation, separating optical and acoustic modes is more important, and the separation persists when the probe width is reduced to 0.5\ps{} (Fig.~\ref{fig:EELS_vs_chunk}(b)). 
Simultaneously, the increased time resolution improves analysis of mode dynamics: intensity changes between windows grow from $\sim10\%$ at $\tau=1\,{\rm ps}$ to $\sim50\%$ at $\tau=0.5\,{\rm ps}$.
For further narrowing of the temporal probe width to 0.25\ps{}, presented in Fig.~\ref{fig:EELS_vs_chunk}(c), we see a smooth, more gradual change in intensity between the time windows, exhibiting up to $\sim75\%$ increase between the first time window and the window starting 1\ps{} after the excitation.
At the same time, however, for such a temporal probe width, the separation between optical and acoustic modes is closing, indicating we are approaching the physical limit of such analysis.

%----------------------------------------%
%---------- SUBSECTION SUMMARY ----------%
%----------------------------------------%

Figure~\ref{fig:EELS_vs_chunk} shows that on a per-case basis, the time resolution of FTE/TACAW can be pushed further, according to the time\textendash{}energy uncertainty principle, showing a relatively quick relaxation event in even more detail.

\section{Summary and Conclusions}

We have presented a computational FTE/TACAW framework for studying non-equilibrium phonon dynamics with simultaneous momentum and energy resolution. 
In this approach, a controlled vibrational excitation is first introduced via frozen trajectory excitation, and the subsequent relaxation is analyzed by performing TACAW on short trajectory windows extracted at different pump\textendash{}probe delays. 
The resulting spectra provide a time-dependent view of energy-resolved phonon redistribution that complements and extends the information available from energy-integrated diffuse-scattering observables.
Both components, FTE and TACAW, could be extended to other quasiparticle excitations, such as magnons, where all of the discussion here should hold.

The framework was demonstrated for three excitation scenarios in two representative materials: face-centered cubic Ni and diamond-cubic SiC. 
These systems were chosen to probe different relaxation pathways, including energy transfer both (1) within and (2) between acoustic branches, as well as (3) between acoustic and optical branches.
A binary compound with chemically distinct sublattices was chosen to host optical and acoustic modes. 
The present results show that the combined FTE/TACAW workflow can follow the time-resolved redistribution of spectral weight in selected regions of $(\vec{q},\omega)$-space during non-equilibrium relaxation. 

Conceptually, the key step is to interpret TACAW as a local spectral analysis of the time-dependent auxiliary exit wavefunction that carries information about the quasiparticles of interest. 
This viewpoint allows one to move beyond the equilibrium setting without directly relying on a global Kubo-type equilibrium construction at each delay time. 
At the same time, we introduce a modified quantum correction term enabling the restoration of part of the information.
Nonetheless, the present non-equilibrium formulation introduces an additional approximation because no simple detailed-balance correction that allows for the treatment of multi- and multiple-phonon phenomena is available for mode-dependent, time-dependent, non-thermal populations. 
The results should therefore be interpreted primarily in terms of transient spectral redistribution and relative spectral evolution.

The present work serves as a first demonstration of time-resolved TACAW in a non-equilibrium setting. 
Future developments should refine the treatment of non-equilibrium quantum corrections, extend the analysis to more complex excitations and materials, and further strengthen the connection to emerging ultrafast electron-scattering experiments with simultaneous temporal, momentum, and energy resolution.

\section*{Acknowledgments}

We acknowledge support from the Swedish Research Council (grant no.\ 2025-04514), the Olle Engkvist Foundation (grant no.\ 214-0331), and the Knut and Alice Wallenberg Foundation (grant no.\ 2022.0079). 
W. M. acknowledges financial support from the Polish National Agency for Academic Exchange under decision BPN/BEK/2022/1/00179/DEC/1. 
The simulations were enabled by resources provided by the National Academic Infrastructure for Supercomputing in Sweden (NAISS) at NSC Centre, partially funded by the Swedish Research Council through grant agreement no. 2022-06725.
We are grateful to Dominik Florian and Aksel Kobiałka for their valuable comments on the manuscript and to Martin O\v{s}mera for his insights and stimulating discussions.

\section*{Author Contributions statement}

\textbf{Wojciech Marciniak}: conceptualization, formal analysis, data curation, investigation, methodology, software, visualization, and writing\,\textemdash\,original draft.
\textbf{Joanna Marciniak}: data curation, investigation, validation, and visualization.
\textbf{Jos\'e \'Angel Castellanos-Reyes}: formal analysis, and methodology.
\textbf{J\'an Rusz}: conceptualization, funding acquisition, project administration, resources, and supervision.
\textbf{All authors} contributed to the review and editing.

\appendix

\section{\label{app:details}Detailed description of calculations procedure}

In this appendix, we describe the computational details beneficial for reproducing the results.
As far as the implementations are concerned, for real\textendash{}reciprocal space transitions in both the excitation and subsequent measurement phase, \fftw{}~\cite{frigo_design_2005} implementation of the fast Fourier transform is used.
The detailed parameter description follows in the next subsections.

\subsection{system geometry details}

The MD simulations were performed in supercells consisting of cubic unit cells, according to space groups 225 ($Fm\bar{3}m$) for fcc-Ni and 216 ($F\bar{4}3m$) for SiC.
It yields the 4-atom conventional fcc unit cell for fcc-Ni, with the single inequivalent atomic site:
\begin{align*}
    \rm Ni: (0,0,0),
\end{align*}
and the 8-atom conventional diamond-cubic unit cell for SiC, with two inequivalent sites:
\begin{align*}
    \rm Si: (0.00,\ 0.00,\ 0.00), \\
    \rm C: (0.25,\ 0.25,\ 0.25),
\end{align*}

The excitation was performed using the corresponding primitive unit cell:
\begin{align*}
    \vec{a}_{1} &= (a/2,\ a/2,\ 0.0), \\
    \vec{a}_{2} &= (a/2,\ 0.0,\ a/2), \\
    \vec{a}_{3} &= (0.0,\ a/2,\ a/2),
\end{align*}
with identical basis sites, yielding a 1-atom primitive unit cell for fcc-Ni and a 2-atom primitive unit cell for 3C-SiC.
We used a 56$\times$56$\times$28 supercell of primitive unit cells for fcc-Ni and a 48$\times$48$\times$24 supercell of primitive unit cells for 3C-SiC, which together uniquely cover the same space as the conventional supercell used for MD.

\subsection{molecular dynamics}

The initial atom velocities corresponding to 600 K were generated from the Maxwell-Boltzmann distribution.
Damping times of 100~\fs{} for the barostat and 1000~\fs{} for the thermostat were assumed, and we began with a 5~\ps{} equilibration before the production phase.
For potentials chosen to describe fcc-Ni and 3C-SiC, we have used the KOKKOS acceleration in \LAMMPS{} to greatly speed up the calculations.

\subsection{excitation}

Excitation is introduced by FTE in the reciprocal space (body-centered-cubic symmetry).
In the present work, the selected excitations are not arbitrarily localized perturbations. 
They are deliberately chosen so that particular harmonic relations, such as $(2\vec{q}_{\rm ex},2\omega_{\rm ex})$ or $(3\vec{q}_{\rm ex},3\omega_{\rm ex})$, fall onto other physically relevant branches or nearby points on the same branch. 
The equilibrium dispersion, therefore, determines whether the proposed transfer channels are plausible in the first place.
The excitation center points are presented in Tab~\ref{tab:excitation-points}.
We provide all parameters in the fractional coordinates of the primitive as well as the cubic cartesian unit cell.

\begin{table}[h]
    \caption{\label{tab:excitation-points}Excitation points coordinates (fractional), and frequency.}
    \centering
    \begin{tabular}{cccc}
        \hline
        excitation & primitive & cartesian & frequency (THz)\\
        \hline
        Ni-$\Delta$ & ($\frac{13}{56}$, $\frac{13}{56}$, 0) & ($\frac{13}{28}$, 0, 0) & 4.15 \\
        SiC-$\Delta$ & ($\frac{1}{8}$, $\frac{1}{8}$, 0) & ($\frac{1}{4}$, 0, 0) & 3.5 \\
        SiC-$\Sigma$ & ($\frac{13}{32}$, $\frac{13}{64}$, $\frac{13}{64}$) & ($\frac{13}{32}$, $\frac{13}{32}$, 0) & 12.25 \\
        \hline
    \end{tabular}
\end{table}

Around all of the specified points, we apply a 20-fold increase in the phonon population using a Tukey-like band-enhance filter described in Ref~\cite{marciniak_fte_2026}.
In each case, the flat top part of the filter spans $d_1 \times d_2 \times d_3 = \frac{1}{100} \times \frac{1}{100} \times \frac{2}{100}$ of the reciprocal primitive unit lattice vectors and $d_\omega = \frac{2.5}{100}$ of the frequency range.
Also in each case, it decays over $\alpha_1 \times \alpha_2 \times \alpha_3 = \frac{4}{100} \times  \frac{4}{100} \times \frac{6}{100}$ of the reciprocal primitive unit lattice vectors and over $\alpha_\omega = \frac{5}{100}$ of the frequency range.

The trajectory during excitation was also additionally filtered in time using a Tukey (cosine-tapered) window, in the form~\cite{bloomfield_fourier_2004}:

\begin{equation}
\label{eqn:window_tukey}
    W[n] = \begin{cases}
        \frac{1}{2}\left[ 1 - \cos(\frac{2 \pi n}{\alpha N}) \right], & \text{if } 0 \leq n \leq \frac{\alpha N}{2} \\
        1, & \text{if } \frac{\alpha N}{2} \leq n \leq \frac{N}{2} \\
        W[N-n], & \text{otherwise,}
    \end{cases}
\end{equation}

where the sample count $N$ in our case equaled the number of frames in a single equilibrium trajectory, and the oscillation decay interval equaled 0.05 at the beginning and end of the trajectory, totaling $\alpha = 0.1$.

\subsection{relaxation}

\begin{figure}[h]
\includegraphics[width=0.9\columnwidth]{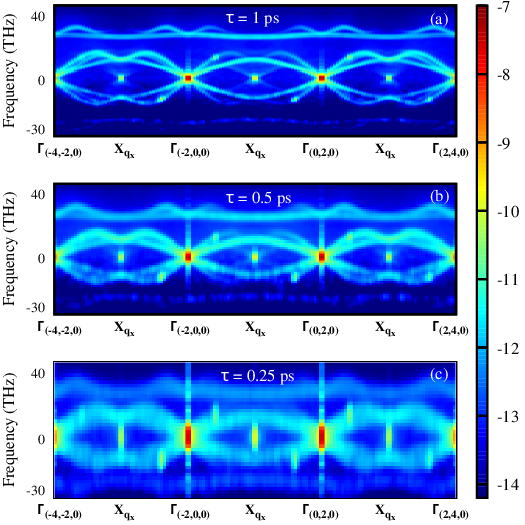}
\caption{\label{fig:spectrum_vs_chunk}EELS/EEGS spectra along the $\Sigma$ high-symmetry line in 3C-SiC, in the time window ending at 1\ps{} into the relaxation.
Panel (a) shows the same data as in Fig.~\ref{fig:spectrum-in-time}(e).
The spectrum with a narrower temporal probe of 0.5\ps{} (corresponding to Fig.\ref{fig:EELS_vs_chunk}(b)) is presented in panel (b).
There, the energy transfer to $(2\vec{q}_{\perp{\rm ex}},2\omega_{\rm ex})$ becomes clearly visible for the intermediate probe temporal width, especially in the energy gain part.
Further decrease of $\tau$, presented in panel (c) (corresponding to Fig.\ref{fig:EELS_vs_chunk}(c)), does not yield more information due to the loss of energy resolution, and an increase of the spectral leakage from the zero-loss peak and its kinematically forbidden reflection.
}
\end{figure}

In the manuscript, we present the transient electron energy-loss and gain spectra during subsequent equilibration.
As we have shown in Sec.~\ref{sec:fte-tacaw-resolution-limits}, reaching a sub-picosecond regime is achievable, though it comes at a cost.
A visual representation of the full $I(\vec{q}_\perp,\omega)$ spectrum along the SiC-$\Sigma$ high-symmetry direction is shown in Fig.~\ref{fig:spectrum_vs_chunk}.
This richer dataset further helps to illustrate the drop in resolution as the time window $\tau$ is reduced.

\subsection{multislice/TACAW}

Finally, in the multislice algorithm, we used a mesh size of 1\,792$\times$1\,792 for fcc-Ni and 1\,536$\times$1\,536 for 3C-SiC.
We used one slice for every atomic plane, totaling 56 slices for fcc-Ni and 96 slices for 3C-SiC,

For efficiency in multislice simulations, we used \verb+pyms+, which is built using the PyTorch library and thus leverages GPU acceleration.
The calculations were compared against the framework employed in Ref.~\cite{marciniak_fte_2026}, employing \drprobe{}.
The comparison showed both results being in the range of numerical error, with no significant qualitative or quantitative differences.

\begin{figure}[h]
\includegraphics[width=0.9\columnwidth]{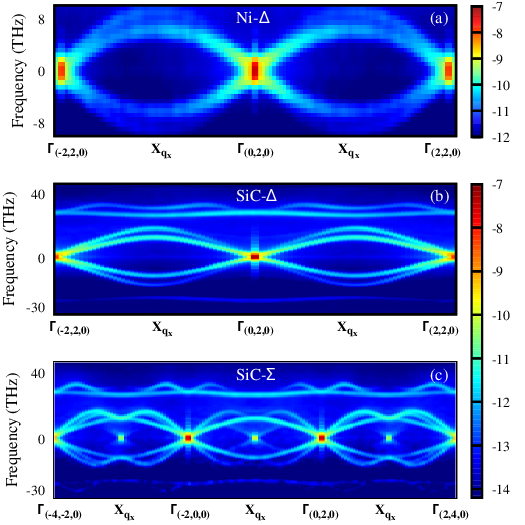}
\caption{\label{fig:spectrum_eq}Equilibrium spectra obtained by averaging over the 5\ps{} long segment of trajectories taken long after the excitation.
%and averaged over 25 realizations of the relaxation.
%
A 25\textendash{}30\ps{} time window was used for the the Ni-$\Delta$ case (a) and SiC-$\Sigma$ case (c), while 145\textendash{}150\ps{} time window was used for SiC-$\Delta$ excitation (b).
}
\end{figure}

Such parameters lead to the equilibrium spectra presented in Fig~\ref{fig:spectrum_eq}.
Those are the spectra of an equilibrated trajectory, taken from 25 to 30\ps{} into the relaxation after the Ni-$\Delta$ and SiC-$\Sigma$ excitations (a,c), and from 145 to 150\ps{} into the relaxation after the SiC-$\Delta$ excitation (b).
Each spectrum is averaged over 25 different trajectories.
We used 100 trajectory frames and an additional 50 matrix elements of zero-padding for the TACAW FFT in both SiC cases and 20 trajectory frames plus 10 additional matrix elements of zero-padding in the case of Ni.
There is no additional overlap among the trajectory fragments used in TACAW, which totals 125 samples for averaging.
This is the intensity used as the equilibrium reference for calculating the transient classical mode temperature during the relaxations, according to Eqs.~\ref{eq:kubo-1} and~\ref{eq:kubo-2}.

\bibliography{fte_tacaw}

\end{document}